\documentclass[prd,amsmath,preprintnumbers,superscriptaddress,nofootinbib]{revtex4}
\usepackage[dvips]{color,graphicx}
\usepackage{dcolumn}
\usepackage{bm}
\usepackage{amsmath}
\usepackage{amssymb}
\newcommand{\be}{\begin{equation}}

\newcommand{\ee}{\end{equation}}
\newcommand{\bea}{\begin{eqnarray}}
\newcommand{\eea}{\end{eqnarray}}

\begin{document}
\title{Neutral quark matter in a Nambu-Jona Lasinio model with vector
interaction}
\author{H. Abuki}\email{abuki@th.physik.uni-frankfurt.de}
\affiliation{Institut f\"ur Theoretische Physik, J.W. Goethe
Universit\"at, D-60438 Frankfurt am Main, Germany}
\author{R. Gatto}\email{raoul.gatto@physics.unige.ch}
\affiliation{D\'epartement de Physique Th\'eorique, Universit\'e de
Gen\`eve, CH-1211 Gen\`eve 4, Suisse}
\author{M. Ruggieri}\email{marco.ruggieri@ba.infn.it}
\affiliation{I.N.F.N., Sezione di Bari, I-70126 Bari, Italia}
\date{\today}
\begin{abstract}
We investigate the three flavor Nambu-Jona Lasinio model of neutral
 quark matter at zero temperature and finite density, keeping into
 account the scalar, the pseudoscalar and the Kobayashi-Maskawa-'t
 Hooft interactions as well as the repulsive vector plus axial-vector
 interaction terms (vector extended NJL, VENJL in the following).
We focus on the effect of the vector interaction on the chiral
 restoration at finite density in neutral matter. We also study the
 evolution of the charged pseudoscalar meson energies as a function of
 the quark chemical potential.
\end{abstract}
 \maketitle \preprint{BA-TH/606-09}

\section{Introduction}
Quantum Chromodynamics (QCD) is accepted nowadays as the theory of strong interactions. The phase diagram of strongly
interacting matter in the $T-\mu$ plane, where $T$, $\mu$ denote respectively the temperature and the baryon chemical
potential, is one of the most intriguing research topics in modern theoretical physics. Investigations on the various
phases of the QCD phase diagram enlighten the physics of high temperature/small density, as well as small
temperature/large density, matter: the former can be created in our laboratories by means of heavy ions collisions; the
latter may be realized in the cores of the compact stellar objects (white dwarfs and neutron stars). The equation of
state of strongly interacting matter is all what one needs to make theoretical predictions in the whole $T-\mu$ regime.
Unfortunately, QCD is analytically treatable only in the perturbative regime. Moreover, the most interesting phenomena
are not of a perturbative nature. For this reason, the QCD equation of state is only known in a small slice of the
phase diagram.

The 
most concrete knowledge about strong interactions at $\mu=0$ comes from Lattice QCD (LQCD). However, at $\mu>0$ LQCD
calculations with three
colors 
suffer from the sign problem. In order to circumvent this problem several approaches have been suggested: expansion in
$\mu/T$ \cite{Allton:2002zi,Gavai:2003mf}, reweighting tecniques \cite{Fodor:2004nz} and analytical continuation from
the imaginary chemical potential axes \cite{de Forcrand:2002ci,D'Elia:2004at}. However none of them has not yet been of
a practical use at high chemical potential and small temperature region. Therefore, to make theoretical investigations
on the QCD phase diagram, and to compute the equation of state in a wider range of chemical potentials where
perturbative QCD does not work,
some effective models are needed.
Among them the Nambu-Jona Lasinio model \cite{%
Nambu:1961tp,Klevansky:1992qe,Hatsuda:1994pi,Buballa:2003qv} is probably
the most popular one: it shares the global symmetries of QCD and is simpler to
handle than QCD itself. The main characteristic of the QCD vacuum, that
is spontaneous breaking of the chiral symmetry, is described in the NJL
model in a 
clear way. Moreover its minimal extension to the Polyakov-NJL
model \cite{Fukushima:2003fw,Ratti:2005jh} reproduces LQCD results
at $\mu=0$ as well as at small chemical potential \cite{Ratti:2007jf}
and at imaginary chemical potential \cite{Sakai:2008py}.
As a consequence NJL model is a promising tool to make calculations of the QCD phase diagram.

In this work, we investigate the three flavor Nambu-Jona Lasinio model of neutral and $\beta$-equilibrated quark matter
at zero temperature and finite density, keeping into account the scalar, the pseudoscalar and the Kobayashi-Maskawa-'t
Hooft interactions \cite{'tHooft:1976up,Kobayashi:1970ji} as well as the repulsive vector plus axial-vector interaction
terms \cite{Bernard:1988db,Asakawa:1989bq,Klimt:1989pm,Klimt:1990ws}. The introduction of the vector interaction and
thus of the vector excitations is interesting for several reasons: firstly, it is well known that vector interactions
play a dominant role in determining the properties of matter at intermediate densities. Experimental progresses on the
measurements of the in-medium properties of vector mesons can be found in Refs.~\cite{:2007mga,Naruki:2005kd}. In these
conditions of baryon densities the main degrees of freedom are nucleons and the lighter pseudoscalar mesons ($\pi$ and
$K$), interacting among them by the exchange of vector mesons (mainly $\rho$ and $\omega$). One may wish to derive
meson-baryon interactions starting from a microscopic interacting theory of the constituent quarks. Such a derivation
can be obtained by means of the well known bosonization and hadronization of the NJL
lagrangian~\cite{Bentz:2001vc,Bentz:2002um}. Keeping in mind the relevance of vector meson modes on the phenomenology
of nuclear matter at intermediate densities, the vector interaction (which excites vector and pseudovector mesons) must
be included from the very beginning in the hadronization of the NJL lagrangian. Secondly, vector meson exchange might
be responsible for kaon condensation at high density. With respect to this phenomenon, it is of a certain interest to
compare the scenario offered by the NJL model with that obtained within the Hidden Local Symmetry framework, the latter
being consistent with kaon condensation~\cite{Harada:1992bu,Harada:2000at,Harada:2003jx,Harada:2000kb}. Thirdly, it has
been suggested that quark hadron continuity can be realized by means of the spectral function continuity of the vector
mesons~\cite{Yamamoto:2007ah}.

These are only a part of the reasons that lead us to consider the role of the vector interaction in quark matter, with
particular focus on  neutral and $\beta-$equilibrated systems which should be realized in the core of neutron stars, if
it exists.

We study in this work the extended NJL model with vector interaction, with particular emphasis put on the neutral and
equilibrated quark matter. The role of the vector interaction in NJL model has been studied several
times~\cite{Bernard:1988db,Asakawa:1989bq,%
Klimt:1989pm,Klimt:1990ws,Lutz:1992dv,%
Kitazawa:2002bc,Fukushima:2008wg}, but not yet in the regime of electrical neutrality and $\beta$ equilibrium. The
study of neutral, as well as $\beta$-equilibrated, matter is interesting for astrophysical applications: matter inside
a compact star has to be electrically neutral; moreover, since weak processes have a small characteristic time compared
to the lifetime of a compact star, matter inside it
should be 
equilibrated with respect to weak-interactions also. We can anticipate one of the results of our study, namely the
phase diagram we compute does not differ qualitatively from that found in the aforementioned references. However, this
does not diminish the relevance of our work. As a matter of fact, if one wishes to study the ground state of the model,
having in mind physical applications as for example the structure of the compact stellar objects, then neutrality is a
necessary requirement that must be settled in, and not a mere academic problem.

The vector coupling $G_V$ in the vacuum can be determined by the fit of the vector meson spectrum. However it is not
clear if $G_V$ in the nuclear medium has the same value it has in the vacuum~\cite{Fukushima:2008wg}. Experimental data
on the medium modifications of the $\omega$ vector meson mass~\cite{Naruki:2005kd} for densities up to the saturation
nuclear density can be reproduced within a NJL model with a fixed value of $G_V$ but with additive multiquark
interactions~\cite{Huguet:2006cm,Huguet:2007jc}; the effect of these interactions can be rephrased simply as density
dependent redefinitions of the coupling constants~\cite{Huguet:2006cm,Huguet:2007jc,Kashiwa:2006rc,Kashiwa:2007pc}. As
a consequence it is not wrong to think to a model in which only four quark interactions are included but where the
coupling constants run with density. Instead of computing $G_V$ at each value of the chemical potential in this work we
treat it as a free parameter, and investigate on its influence on the restoration of the approximate chiral symmetry at
finite density. NJL studies with $G_V = 0$ predict a first order chiral restoration at $\mu$ of the order of the
constituent quark masses. When $G_V$ is switched on, if its magnitude is larger than a critical value, then the
transition becomes a smooth crossover~\cite{Klimt:1990ws}. We reproduce this scenario with neutral matter at
equilibrium. Also, a strong enough vector interaction disfavors the existence of the critical end point of the phase
diagram~\cite{Fukushima:2008wg}.

We do not include color superconductivity in this study. This is done
for simplicity. Nevertheless our results for the electron chemical
potential and for the densities of the various species 
bear some implication of the effects of
the repulsive vector interaction 
in the color superconductive phases.
A quantitative study on the role of vector (as well as multiquark)
interaction on the two flavor color superconductor has been performed in
Refs.~\cite{Kitazawa:2002bc,Kashiwa:2007pc}, but an analogous study in
realistic three flavor quark matter is still missing.

Another point that we consider in some detail is the spectrum of the pseudoscalar excitations. This subject has been
considered, within the NJL model but in the non-neutral case, in Ref.~\cite{Lutz:1992uw}. We firstly derive the loop
expansion of the meson action. It includes scalar, pseudoscalar, vector and pseudovector mesons. However we focus here
only on the pseudoscalar modes, therefore we specialize our equations to this case. We leave a more complete study to a
next work. This choice is motivated by the comparison we would be tempted to make with nuclear matter models, which
predict a kaon condensed phase for densities in the range $(2.5-5)\rho_0$ \cite{Kaplan:1986yq}. We anticipate another
result, namely the absence of strangeness condensation even in presence of a vector interaction, although the vector
interaction lowers the $K^{-}$ in-medium energy. This result was anticipated in Ref.~\cite{Lutz:1992uw}. However, the
authors of this paper do not make any attempt to compute the chemical potential felt by the kaons, which is a relevant
ingredient to exclude definitely meson condensation in the ground state, since they are not interested to the neutral
phases. Moreover, they made their analysis on the basis of the low density approximation within the lowest linear level
in $\rho/\rho_0$. Even though at this order some model independent prediction is possible, one might be rather
interested in the outcome of the model without any further approximations other than the random phase approximation
(RPA) together with the mean field approximation (MFA).

With this in mind, our work improves, and at the same time confirms, the results of Ref.~\cite{Lutz:1992uw}, since we
compute on the same footing both the meson masses and their effective chemical potentials, in the neutral phases.
Moreover, at large enough density we find a new collective mode with $K^-$ quantum number but with lower mass, standing
just below the Landau damping threshold. This is an unexpected feature of the model, which comes out after the low
density approximation is relaxed. Even if we cannot exclude the possibility that this new excitation it is just a model
artifact, it should be stressed that it might modify the arguments about the critical baryon density for the onset of
strangeness, and therefore it might even bring a drastic change in the conventional picture for the population of
strangeness in the compact stars.

We also compare the NJL scenario studied here with that of a simple nuclear model in which kaon condensation is
realized~\cite{Glendenning:1997wn,Prakash:1996xs}; this comparison is useful since it suggests which are the possible
directions to follow in order to reproduce, at least qualitatively, the scenario accepted by the nuclear matter physics
community. For completeness we notice that other studies exist in which the kaon condensation scenario is not
favored~\cite{Roth:2005nx}.

The plan of the paper is as follows: in section II we specify the lagrangian of the vector extended NJL model. In
section III we derive the mean field effective action of the VENJL model as well as the effective action for the meson
excitations at the second order in the meson fields. In section IV we show results for the restoration of the
approximate chiral symmetry at finite baryon chemical potential. In section V we present a detailed discussion of the
charged pseudoscalar modes in neutral and $\beta-$equilibrated quark matter. In section VI, inspired by our results, we
briefly discuss on the role that vector interaction might have on color superconductivity. Finally in section VII we
summarize our results and draw our conclusions.

\section{The VENJL Model}
In this work we are interested in neutral and $\beta$-equilibrated quark matter at finite density. We work in the grand
canonical ensemble formalism, introducing a chemical potential $\mu$ corresponding to the conserved baryon number. To
be more specific we consider a system of $u$, $d$ and $s$ quarks at a finite chemical potential $\mu$ described by the
lagrangian
\begin{eqnarray}
{\cal L} &=& \sum_f\bar\psi_f\,\left[i\partial_\mu \gamma^\mu  + \mu\gamma_0\right]\psi_f + {\cal L}_{mass} \nonumber
\\
&&~~~~~+~{\cal L}_4~+~{\cal L}_6~+~{\cal L}_V~+~{\cal L}_A \nonumber \\
&&~~~~~+~ \bar{e}\,\left[i\partial_\mu \gamma^\mu \right]e ~,\label{eq:lagr1}
\end{eqnarray}
where $\psi_f$ corresponds the quark field with flavor $f$ $(=1,2,3$ for $u,d,s$) and $e$ denotes the electron field.
We now specify each term in Eq.~\eqref{eq:lagr1}. A sum over color indices is understood in Eq.~\eqref{eq:lagr1}.

Equilibrium under weak interactions $d\rightarrow u e \bar{\nu}$, $s\rightarrow u e \bar{\nu}$ implies $\mu_d = \mu_s$
and $\mu_d = \mu_u + \mu_e$ (we assume neutrinos escape from matter
therefore $\mu_\nu = 0$\footnote{%
Some interesting possibilities may also arise when the neutrinos are trapped in the very early stage of the thermal
evolution of proto-neutron stars before the deleptonization. The readers are reffered to
\cite{Ruester:2005ib,Sandin:2007zr,Abuki:2009hx,Laporta:2005be} in this context.}). Moreover, in order to achieve
electrical neutrality we introduce the Lagrange multiplier $\mu_Q$ associated to the total charge $Q$, add the term
$\mu_Q \hat{Q}$ to the lagrangian~\eqref{eq:lagr1} and minimize the thermodynamic potential under the stationarity
condition $\partial\Omega/\partial\mu_Q = 0$. The total charge operator is given by
\begin{equation}
\hat{Q} = \frac{2}{3}u^\dagger u - \frac{1}{3}d^\dagger d - \frac{1}{3}s^\dagger s - e^\dagger e~,
\end{equation}
thus adding $\mu_Q \hat{Q}$ to Eq.~\eqref{eq:lagr1} and recognizing that $\mu_Q = -\mu_e$ we get the chemical
potentials of the quarks, namely
\begin{equation}
\mu_u = \mu -\frac{2}{3}\mu_e~,~~~~~\mu_d = \mu_s = \mu +\frac{1}{3}\mu_e~.\label{eq:mus1}
\end{equation}

Next we describe the other terms in our lagrangian~\eqref{eq:lagr1}. The mass term is \be{\cal L}_{mass} ~=~-\,
\sum_fm_f\bar\psi_f\psi_f \ee and $m_f$ is the current mass. In this work we assume $m_u = m_d$. The NJL four-fermion
and six-fermion interaction Lagrangians are~\cite{Klevansky:1992qe,Hatsuda:1994pi,Buballa:2003qv} \bea {\cal L}_4 &=&
G\sum_{a=0}^8\left[\left(\bar\psi \lambda_a \psi\right)^2 + \left(i\bar\psi \gamma_5\lambda_a \psi\right)^2
\right]\label{eq:full4}~,\\ {\cal L}_6 &=& -K\left[{\rm det}\bar\psi_f(1+ \gamma_5)\psi_{f'} \,+\,{\rm
det}\bar\psi_f(1- \gamma_5)\psi_{f'} \right] ~,\label{eq:full6} \eea where $\lambda_a$ are the Gell-Mann matrices in
flavor space ($\lambda_0 = \sqrt{2/3}~{\bm 1}_f$) and the determinant is in flavor space as well.

Finally ${\cal L}_V + {\cal L}_A$ denote the following $U(3)_V\otimes U(3)_A$ invariant interaction  term:
\begin{equation}
{\cal L}_V + {\cal L}_A = - G_V\sum_{a=0}^8 \left[(\bar\psi \gamma^\mu \lambda_a \psi)^2 + (\bar\psi \gamma^\mu
\gamma_5\lambda_a \psi)^2 \right]~, \label{eq:Lva}
\end{equation}
where a summation over color and flavor is understood. In the above equation $\lambda_a$ denote the same set of
matrices in the flavor space introduced in Eq.~\eqref{eq:lagr1}. Typical
values of $G_V$ in the vacuum are $0.2 \le G_V/G \le
3$~\cite{Bernard:1988db,Klimt:1990ws,Lutz:1992dv,Kashiwa:2007pc,Huguet:2006cm}.

\section{Effective potential and effective action of mesons}
In this section we sketch the derivation of the mean field effective potential as well as of the effective action of
the meson modes in the VENJL model. The derivation is done in some detail in order to easily compare our notations and
results with the existing literature. We follow the conventions of Ref.~\cite{Klevansky:1992qe}.

The easiest way to derive the meson propagators is the so called linear
approach~\cite{Belkov:1993fu,Ruiz Arriola:1995ea,Bernard:1995hm}.
Firstly we write the t'Hooft term as an effective 4-fermion interaction:
this is a well known procedure~\cite{Klevansky:1992qe} therefore we do
not insist on its details. After this is achieved the scalar and
pseudoscalar interaction is written as
\begin{equation}
{\cal L}_4 + {\cal L}_6 = \sum_{a=0}^8\left[K_a^{(-)}(\bar\psi \lambda_a \psi)^2 + K_a^{(+)}(\bar\psi
i\gamma_5\lambda_a \psi)^2\right] + {\cal L}_{mixing}~,\label{eq:OOOfff}
\end{equation}
where the term ${\cal L}_{mixing}$ generates the mixing between $\pi^0$, $\eta^0$ and $\eta^8$ modes and is not
important in this context since we focus on the charged meson modes. The effective coupling constants are defined as
follows:
\begin{eqnarray}
K_0^{(\pm)} &=& G \mp\frac{K}{3}(-\sigma_s -\sigma_u -\sigma_d)~, \\
K_1^{(\pm)} &=& K_2^{(\pm)} = K_3^{(\pm)} =G \pm\frac{K}{2}(-\sigma_s )~, \\
K_4^{(\pm)} &=& K_5^{(\pm)} = G \pm\frac{K}{2}(-\sigma_d )~, \\
K_6^{(\pm)} &=& K_7^{(\pm)} = G \pm\frac{K}{2}(-\sigma_u )~, \\
K_8^{(\pm)} &=&G \mp \frac{K}{6}(-\sigma_s +2\sigma_u +2\sigma_d)~.
\end{eqnarray}
In the above equations $\sigma_f = \langle\bar f f\rangle$. Then we define the following meson fields:
\begin{eqnarray}
\phi_a &=& K_a^{(-)} \bar\psi \lambda_a \psi~,  \\
\pi_a &=& K_a^{(+)} \bar\psi i \gamma_5 \lambda_a \psi~, \\
v_a^\mu &=& G_V \bar\psi \gamma^\mu \lambda_a \psi~, \\
a_a^\mu &=& G_V \bar\psi \gamma^\mu \gamma_5 \lambda_a \psi~.
\end{eqnarray}
The partition function of the model can be cast in the form~\cite{Bernard:1995hm}
\begin{equation}
{\cal Z} = \int {\cal D}M {\cal D}\psi {\cal D}\bar\psi e^{i S}~,~~~~~{\cal D}M \equiv {\cal D}\phi{\cal D}\pi{\cal
D}v{\cal D}a~,
\end{equation}
with the bosonized action given by
\begin{eqnarray}
S &=& \int d^4 x~\bar\psi \left[i\gamma^\mu\partial - m +2 \phi_a \bar\psi \lambda_a \psi +2 \pi_a \bar\psi i
\gamma_5\lambda_a \psi -2 v_a^\mu \bar\psi \gamma_\mu \lambda_a \psi -2 a_a^\mu \bar\psi \gamma_\mu \gamma_5\lambda_a
\psi \right]\psi \nonumber \\
&& - \int d^4 x~\left[\frac{\phi_a^2}{K_a^{(-)}} +  \frac{\pi_a^2}{K_a^{(+)}} \right] + \int d^4 x~\left[\frac{v_a^2 +
a_a^2}{G_V} \right]~. \label{eq:coco}
\end{eqnarray}
In Eq.~\eqref{eq:coco} we have not shown explicitly the contribution of the free electron gas (it will be inserted at
the end of the calculation). At this stage the meson fields are external fields (i.e. with no kinetic term). Their
kinetic terms as well as their interactions will arise once the fermions are integrated out. Before performing this
integral we notice that in our application we expect condensation in some of the $\phi_a$ and $v^\mu_a$ channels. These
condensations are related to chiral condensates of the three flavors and to fermion densities. Thus we define
\begin{eqnarray}
\phi_a &=& \sigma_a + \delta_a~, \label{eq:30}\\
v^\mu_a &=& V^\mu_a + \rho^\mu_a~,\label{eq:31}
\end{eqnarray}
with $a=0,\dots,8$ and $\langle\delta_a\rangle = \langle\rho^\mu_a\rangle = 0$. With this choice ${\cal D}M = {\cal
D}\delta{\cal D}\pi{\cal D}\rho{\cal D}a$ and the physical meson fields will be identified with the fluctuations around
their expectation value. In principle we should introduce expectation values for charged pseudoscalar mesons as well,
since their in-medium energies can be lower than the threshold of
condensation.
However we firstly study the system without the assumption of charged
pseudoscalar meson condensation: as long as we exclude the possibility
of first order transition \cite{Abuki:2008wm,He:2005nk}, the evolution
of their rest energies as a function of the baryon density will allow to
establish wether condensation occurs or not (see next section for more
details).

The functional integration over ${\cal D}\psi {\cal D}\bar\psi$ in the partition function can be done exactly since the
action~\eqref{eq:coco} is quadratic in the fermion fields. Using textbook relations we get
\begin{equation}
{\cal Z} = \int {\cal D}M e^{i S[\phi,\pi,v,a]}~, \label{eq:Z}
\end{equation}
with the effective action $S[\phi,\pi,v,a]$ given by
\begin{eqnarray}
S[\phi,\pi,v,a] &=& -i\text{Tr}\log\left[S_0^{-1} + 2\delta_a \lambda_a + 2i\pi_a \gamma_5 \lambda_a -2\rho^\mu_a
\gamma_\mu \lambda_a -2a^\mu_a \gamma_\mu \gamma_5\lambda_a\right] \nonumber \\
&& - \int d^4 x~\left[\frac{\sigma_a^2 + \delta_a^2 +2\delta_a \sigma_a}{K_a^{(-)}} +  \frac{\pi_a^2}{K_a^{(+)}}
\right] \nonumber \\
&&+ \int d^4 x~\left[\frac{V_a^2  + \rho_a^2 +2 V_a\rho_a + a_a^2}{G_V} \right]~,
\end{eqnarray}
and
\begin{equation}
S_0^{-1} = i\gamma_\mu\partial^\mu - m +\mu\gamma_0 +2\sigma_a \lambda_a -2 V^\mu_a \gamma_\mu \lambda_a~.
\end{equation}
For seek of compactness we introduce the total fluctuating meson field
\begin{equation}
{\cal M}_a = \delta_a  + i\pi_a \gamma_5  -\rho^\mu_a \gamma_\mu  -a^\mu_a \gamma_\mu \gamma_5~.
\end{equation}
Expanding the $\log$ term we finally cast the effective action in the form
\begin{eqnarray}
S[\phi,\pi,v,a] &=& S_{MF}-i \sum_{n=1}^\infty\frac{(-1)^{n+1}}{n}\text{Tr} \left[(S_0 2\lambda_a {\cal M}_a)^n\right]\nonumber \\
&& - \int d^4 x~\left[\frac{\delta_a^2 +2\delta_a \sigma_a}{K_a^{(-)}} +  \frac{\pi_a^2}{K_a^{(+)}}
\right] \nonumber \\
&&+ \int d^4 x~\left[\frac{\rho_a^2 +2 V_a\rho_a + a_a^2}{G_V} \right]~,\label{eq:APP1}
\end{eqnarray}
where $S_{MF}$ denotes the contribution of the effective action that does not depend on the fluctuating fields,
\begin{equation}
S_{MF} = -i\text{Tr}\log S_0^{-1} - \int d^4 x~\left[\frac{\sigma_a^2 }{K_a^{(-)}} \right] + \int d^4
x~\left[\frac{V_a^2 }{G_V} \right]~.\label{eq:SMF}
\end{equation}
The sum in Eq.~\eqref{eq:APP1} is a loop expansion that generates the kinetic terms for the mesons as well as their
interactions. We now examine $S_{MF}$ and this loop expansion separately.

\subsection{The mean field effective potential}
We now focus on the mean field term in Eq.~\eqref{eq:SMF}. Firstly we have to specify the expectation values in
Eqs.~\eqref{eq:30} and~\eqref{eq:31}. We assume that in the ground state condensation occurs only in the channels
$\langle\bar{f}f\rangle$ and $\langle f^\dagger f\rangle$ (the latter being relevant at finite quark density). Here $f$
denotes the quark with flavor $f$. This implies that in Eqs.~\eqref{eq:30} and~\eqref{eq:31} only the terms with
$a=0,3,8$ and $\mu=0$ survive, the reminders being zero. It is easy to show that $S^{-1}_0$ takes the simple form
\begin{equation}
S^{-1}_0 = \left(\begin{array}{ccc}
  h_u & 0 & 0 \\
  0 & h_d & 0 \\
  0 & 0 & h_s \\
\end{array}\right)
\end{equation}
where
\begin{equation}
h_f = (p_0 + \mu_f -4 G_V \rho_f)\gamma_0 -\bm{p}\cdot\bm{\gamma} - M_f~, \label{eq:HpropFF}
\end{equation}
and
\begin{equation}
M_f = m_f - 4 G \sigma_f~+2\,\,K\,\sigma_{f+1}\,\sigma_{f+2}\ \label{eq:masseMU0}
\end{equation}
denotes the mean field (or constituent) quark mass. In the above equation
\begin{equation}
\sigma_f =  -i N_c\, \text{tr}S_f ~,
\end{equation}
denotes the chiral condensate of the flavor $f$, where $S_f$ is the propagator of the quark of flavor $f$, $N_c$ is the
number of colors, and the trace is on spinor indices only. Also we have defined
$\sigma_4=\sigma_u,\,\sigma_5=\sigma_d$. Analogously $\rho_f = \langle f^\dagger f\rangle$ denotes the number density
of $f$.

From Eq.~\eqref{eq:HpropFF} the role of the vector interaction in the mean field approximation is clear: it shifts the
quark chemical potentials~\eqref{eq:mus1} to
\begin{equation}
\mu_f \rightarrow \mu_f -4 G_V \rho_f~. \label{eq:shift}
\end{equation}
Thus depending on the sign of $G_V$ the chemical potentials will be shifted either upwards or downwards once the number
density $\rho_f \neq 0$.

We notice that the flavor channels $\lambda_0$, $\lambda_3$ and $\lambda_8$ in the vector interaction generate coupling
of the quark currents to the total quark number density $\rho_u + \rho_d +\rho_s$, the isospin density $\rho_u -
\rho_d$ and to the hypercharge density $\rho_u + \rho_d - 2\rho_s$
respectively. This is similar to what happens in the relativistic mean
field nuclear models~\cite{Glendenning:1997wn}, where the coupling of
the isospin nuclear current to the
isovector $\bm\rho^\mu = (\rho^\mu_1,\rho^\mu_2,\rho^\mu_3)$ and of the
nuclear current to the vector $\omega^\mu$
meson gives rise at the mean field level to expectation values of the zeroth component of the $\omega^\mu$ and
$\rho^\mu_3$ fields, which result proportional respectively to the total baryon and to the isospin densities. We simply
mention here that one cannot identify the vectors of the channels $\lambda_0$, $\lambda_3$ and $\lambda_8$ with the
physical $\omega$, $\rho_0$ and $\phi$ fields since a mixing among them occurs in the vacuum as well as at finite
density. Keeping this in mind, with an abuse of notation we define
\begin{eqnarray}
&&\langle\omega^0\rangle = G_V\langle u^\dagger  u + d^\dagger d + s^\dagger s\rangle = G_V(\rho_u + \rho_d + \rho_s)~, \label{eq:omD}\\
&&\langle\rho^0_3\rangle = G_V\langle u^\dagger  u - d^\dagger d \rangle = G_V(\rho_u - \rho_d)~, \label{eq:omE}\\
&&\langle\phi^0\rangle = G_V\langle u^\dagger  u + d^\dagger d -2 s^\dagger s\rangle = G_V(\rho_u + \rho_d -2
\rho_s)~,\label{eq:omF}
\end{eqnarray}
or equivalently
\begin{eqnarray}
\rho_u &=& \frac{1}{6G_V}\left(2\langle\omega^0\rangle + 3\langle\rho^0_3\rangle + \langle\phi^0\rangle\right)~, \label{eq:rhoU}\\
\rho_d &=& \frac{1}{6G_V}\left(2\langle\omega^0\rangle - 3\langle\rho^0_3\rangle + \langle\phi^0\rangle\right)~, \label{eq:rhoD}\\
\rho_s &=& \frac{1}{6G_V}\left(2\langle\omega^0\rangle -2 \langle\phi^0\rangle\right)~. \label{eq:rhoS}
\end{eqnarray}
The above equations allow to rephrase the effective chemical potential in Eq.~\eqref{eq:shift} in terms of the
expectation values of the vector meson fields.

Finally writing
\begin{equation}
e^{-\beta V \Omega} = {\cal Z}
\end{equation}
with ${\cal Z}$ given in Eq.~\eqref{eq:Z} we get the mean field effective potential:
\begin{equation}
\Omega_{MF} = -4K\sigma_u\sigma_d\sigma_s+2G\!\sum_{f=u,d,s}\sigma_f^2 - 2G_V\!\sum_{f=u,d,s}\rho_f^2 + V_{\log} ~,
\label{eq:OMEGA}
\end{equation}
with
\begin{equation}
V_{\log} = -T\sum_{n=-\infty}^{+\infty}\int\frac{d^3p}{(2\pi)^3}\text{Tr}\log \left[\frac{1}{T}\left(\begin{array}{ccc}
  h_u & 0 & 0 \\
  0 & h_d & 0 \\
  0 & 0 & h_s \\
\end{array}\right)\right]~. \label{eq:chi}
\end{equation}
In the above equations we have introduced a finite temperature $T$ in order to easily handle infrared divergencies that
arise when the chemical potential of the flavor $f$ is larger than its
mean field mass $M_f$. At the end of the calculations we make the limit
$T\rightarrow 0^+$.

The ground state of the model at hand is characterized by the numerical
values of the condensates and of the quark number densities. We
determine the values of the chiral condensates by looking for the global
minima of the total effective potential:
\begin{equation}
\Omega = \Omega_{MF} - \frac{\mu_e^4}{12\pi^2}~,\label{eq:totOmega}
\end{equation}
where the second addendum on the r.h.s is the thermodynamic potential of the free electron gas that will neutralize the
net quark charge. At a fixed value of $\mu$ we choose the value $\bar\mu_e$ of $\mu_e$ which neutralizes the system,
which is defined by the stationarity condition
\begin{equation}
0 = \left.\frac{\partial\Omega}{\partial \mu_e}\right|_{\mu_e = \bar\mu_e}~.\label{eq:neutra}
\end{equation}
The complete numerical strategy will be described in the next section.

\subsection{Effective action of meson fluctuations at the second order}
We now discuss the action of meson fluctuations in Eq.~\eqref{eq:APP1}. The term with $n=1$, when summed to the linear
terms in the second and third line of the same equation, gives rise to terms of the form $\delta_a F_a$ or $\rho_a
G_a$, where $F_a$ and $G_a$ schematically denote the gap equations whose solutions determine the physical values of the
chiral condensate and the quark number densities~\cite{Bernard:1995hm}. Since for these values both $F_a$ and $G_a$
vanish, the linear terms do not appear in the effective action.

Next we consider the term with $n=2$. This gives rise to the kinetic terms of the meson fields as well as to the mixing
of some of the modes. Since the linear terms vanish we can write the effective action at the second order as
\begin{eqnarray}
S[\delta,\pi,\rho,a] &=& 2i \int d^4x~d^4y~\text{Tr} \left[S_0(x,y) \lambda_a {\cal M}_a(y)
S_0(y,x) \lambda_b {\cal M}_b(x)\right]\nonumber \\
&& - \int d^4 x~\left[\frac{\delta_a(x)^2 }{K_a^{(-)}} +  \frac{\pi_a(x)^2}{K_a^{(+)}} \right] + \int d^4
x~\left[\frac{\rho_a(x)^2  + a_a(x)^2}{G_V} \right]~,\label{eq:APP2}
\end{eqnarray}
where we have not written explicitly the terms proportional to $\sigma^2_a$, $V_a^2$ since they contribute only to
$S_{MF}$. In the above equation the trace is understood over color, flavor and Dirac indices.

By means of the loop expansion in Eq.~\eqref{eq:APP2} we could study every meson fluctuation at the second order in the
meson field about the mean field solution. However, in this paper we focus our attention on the charged pseudoscalar
modes, 
reserving a more complete study to a next paper.

The propagator of these modes is easily obtained from Eq.~\eqref{eq:APP2}. We are interested in the rest energies of
the meson modes. We have verified by a direct calculation that there is no mixing between the pseudoscalar and the
longitudinal component of the pseudovector mode in the case we put the spatial momentum of the meson $\bm Q = 0$.
Moreover parity conservation in strong interactions is enough to ensure that mixing at the second order in the fields
does not arise between pseudoscalar mesons and the vector or the scalar
ones.  Thus the remaining term to be computed is
that of order ${\cal O}(\pi_a \pi_b)$ of Eq.~\eqref{eq:APP2}. Fourier transforming both the fields and the propagators
we have
\begin{equation}
S[\pi_a] = -\int\frac{d^4q}{(2\pi)^4} \pi_a(-q) \pi_b(q)\left[\frac{\delta_{ab}}{K_b^{(+)}}
+2i\int\frac{d^4\ell}{(2\pi)^4} \text{Tr}\left(S_0(\ell)\gamma_5 \lambda_a S_0(\ell+ q)\gamma_5 \lambda_b\right)
\right]~. \label{eq:l}
\end{equation}
Since at the second order there is no mixing among the pseudoscalar and the other fields, at least in the rest system
of the mesons, we identify the matrix in parenthesis in Eq.~\eqref{eq:l} with half of the inverse meson propagator in
momentum space:
\begin{equation}
\frac{1}{2}D_{ab}(q)^{-1} = - \frac{\delta_{ab}}{K_b^{(+)}}  - 2i\int\frac{d^4\ell}{(2\pi)^4}
\text{Tr}\left[S_0(\ell)\gamma_5 \lambda_a S_0(\ell+ q)\gamma_5 \lambda_b\right]~.
\end{equation}

Consider for example the charged kaons whose fields are defined as
\begin{equation}
K^{\pm} = \frac{\pi_4 \pm i \pi_5}{\sqrt{2}}~,
\end{equation}
then using $D_{44}^{-1} = D_{55}^{-1}\equiv D$
and $D_{45}^{-1}=-D_{54}\equiv i\delta D^{-1}$ it is easy to show that
\begin{equation}
\begin{array}{l}
\int\frac{d^4q}{(2\pi)^4} \frac{1}{2}%
 (\pi_4(-q,),\pi_5(-q))%
 \left(\begin{array}{cc}
 D^{-1}(q) & i\delta D(q)\\
 -i\delta D(q) & D^{-1}(q)\\
 \end{array}
 \right)\left(\begin{array}{c}
 \pi_4(q)\\
 \pi_5(q)\\
 \end{array}
 \right)\\[2ex]
 =\int\frac{d^4q}{(2\pi)^4}K^{+}(-q)(D^{-1}(q)-\delta D^{-1}(q))K^-(q)
 + K^{-}(-q)(D^{-1}(q)+\delta D^{-1}(q))K^+(q)~;
\end{array}
 \nonumber
\end{equation}
the condition $D^{-1}(q_0 = \omega_K,\bm q = 0)-\delta D^{-1}(q_0=\omega_K,\bm q = 0)$ = 0 with $\omega_K$ being
positive defines the rest energy of charged $K^+$. Since $D^{-1}(-q_0,\bm q )=D^{-1}(q_0,\bm q)$ and $\delta
D^{-1}(-q_0,\bm q)=-\delta D^{-1}(q_0,\bm q)$ can be easily shown, it follows that the negative energy solution to
$D^{-1}-\delta D^{-1}=0$ corresponds to the positive energy solution to $D^{-1}+\delta D^{-1}=0$ indicating that we
need only one equation $D^{-1}-\delta D^{-1}=0$ to extract both the $K^-$ and $K^+$ energies. Similar results hold for
the remaining mesons.

\section{The neutral ground state of the VENJL model}
In this section we discuss the neutral and $\beta$-equilibrated  ground state of the VENJL model at zero temperature,
obtained by the procedure of minimization of the effective potential $\Omega$ under the neutrality
condition~\eqref{eq:neutra}. We present results for the following set of parameters~\cite{Ruester:2005jc}:
\begin{eqnarray}
&&m_u = m_d = 5.5~\text{MeV}~, \label{eq:para1}\\
&&m_s = 140.7~\text{MeV}~, \\
&&\Lambda = 602.3~\text{MeV}~,\\
&& G = 1.835/\Lambda^2~,\\
&& K = 12.36/\Lambda^5~,\label{eq:para2}
\end{eqnarray}
but the results are qualitatively similar to those that we obtain with the different set of
Ref.~\cite{Ippolito:2007uz}. Firstly we are interested in the effect of
the vector interaction on the neutral ground state of the model.

The value of $G_V$ can (and should) be fixed {\em in the vacuum} in order to reproduce the spectrum of the vector
mesons, see for example~\cite{Klimt:1989pm}. However, it is not clear
how large is the modification of $G_V$ in the
medium~\cite{Fukushima:2008wg}, therefore instead to fix it by the meson
spectrum in the vacuum we treat it as a free parameter in order to grasp
its effects on the chiral restoration transition at finite density, as
well as on the in-medium meson properties. We measure the strength of
the vector interaction in terms of the scalar coupling $G$ and introduce
the ratio
\begin{equation}
r_V = \frac{G_V}{G}~.
\end{equation}
In the vacuum we find $r_V \approx 0.6$ in order to reproduce the
neutral $\rho$ meson mass $m_\rho = 775$ MeV. We notice that because of
the shift in Eq.~\eqref{eq:HpropFF} the quark number density for the
flavor $f$ in the mean field approximation is self-consistently defined
by the equation
\begin{equation}
\rho_f = \frac{N_c}{3\pi^2}\left[(\mu - Q \mu_e - 4 G_V \rho_f)^2 -
                M_f^2\right]^{3/2} \theta\left[(\mu - Q
                \mu_e - 4
G_V \rho_f)^2 - M_f^2\right]~,\label{eq:num}
\end{equation}
where $M_f$ denotes the in-medium quark mass. The r.h.s. of the above equation is nothing but the number density of a
degenerate Fermi gas with a density dependent Fermi momentum given by
\begin{displaymath}
\left[(\mu - Q \mu_e - 4 G_V \rho_f)^2 - M_f^2\right]^{1/2}~.
\end{displaymath}
It is interesting to note that Eq.~(\ref{eq:num}) is equivalent to the partial-derivative of the effective potential
with respect to $\rho_f$ which is related to condensations of $\omega$, $\rho_0$ and $\phi$ through
Eqs.~(\ref{eq:omD}-\ref{eq:omF}) {i.e.,}
\begin{equation}
 \frac{\partial\Omega_{MF}}{\partial\rho_f}=0.
\end{equation}
In this case $\Omega_{MF}$ should be regarded as a function of eight variables, i.e., $\{\sigma_f,\rho_f,\mu_e,\mu\}$
\cite{Kitazawa:2002bc}. It turns out that the ground state is realized as the maximum with respect to variables
$\rho_f$ when $G_V>0$ (repulsive vector interaction). However it does not immediately mean the ground state is unstable
against the vector meson fluctuation. Actually the vector meson mode is not tachyonic as we always find a positive and
real mass for $\rho$ meson. This means the curvature of the effective potential cannot help us to judge if the system
is unstable against small fluctuations in the vector channel. The total charge is given by
\begin{equation}
0 = \frac{2}{3}\rho_u -  \frac{1}{3}\rho_d -  \frac{1}{3}\rho_s - \rho_e~,\label{eq:totQ1}
\end{equation}
with quark number densities identified with the solutions of Eq.~\eqref{eq:num}.

Since the number densities in the case $r_V \neq 0$ have to be computed self-consistently by virtue of
Eq.~\eqref{eq:num} the numerical calculations are slightly more involved than for the case $r_V = 0$. Our numerical
strategy is as follows: for each value of $\mu$ we solve Eq.~\eqref{eq:num} for each flavor; this allows to define
three numerical functions $\rho_f = \rho_f(\mu_e,M_f)$, one for each flavor $f$. Moreover we define a fourth numerical
function which relates the physical electron chemical potential (namely the value of $\mu_e$ that corresponds to a
vanishing total electric charge) to the in-medium quark masses. This is achieved by means of the total charge
Eq.~\eqref{eq:totQ1} and densities defined above. We insert the four functions into the effective potential, that now
depends only on the in-medium masses (or equivalently on the chiral condensates). The global minimum of the effective
potential in the space $(M_u, M_d, M_s)$ gives the physical values of the in-medium masses. Once these are known we can
step backward and compute the numerical value of $\mu_e$ and of the number densities.

\begin{figure}[t!]
\centerline{
\includegraphics[width=8cm]{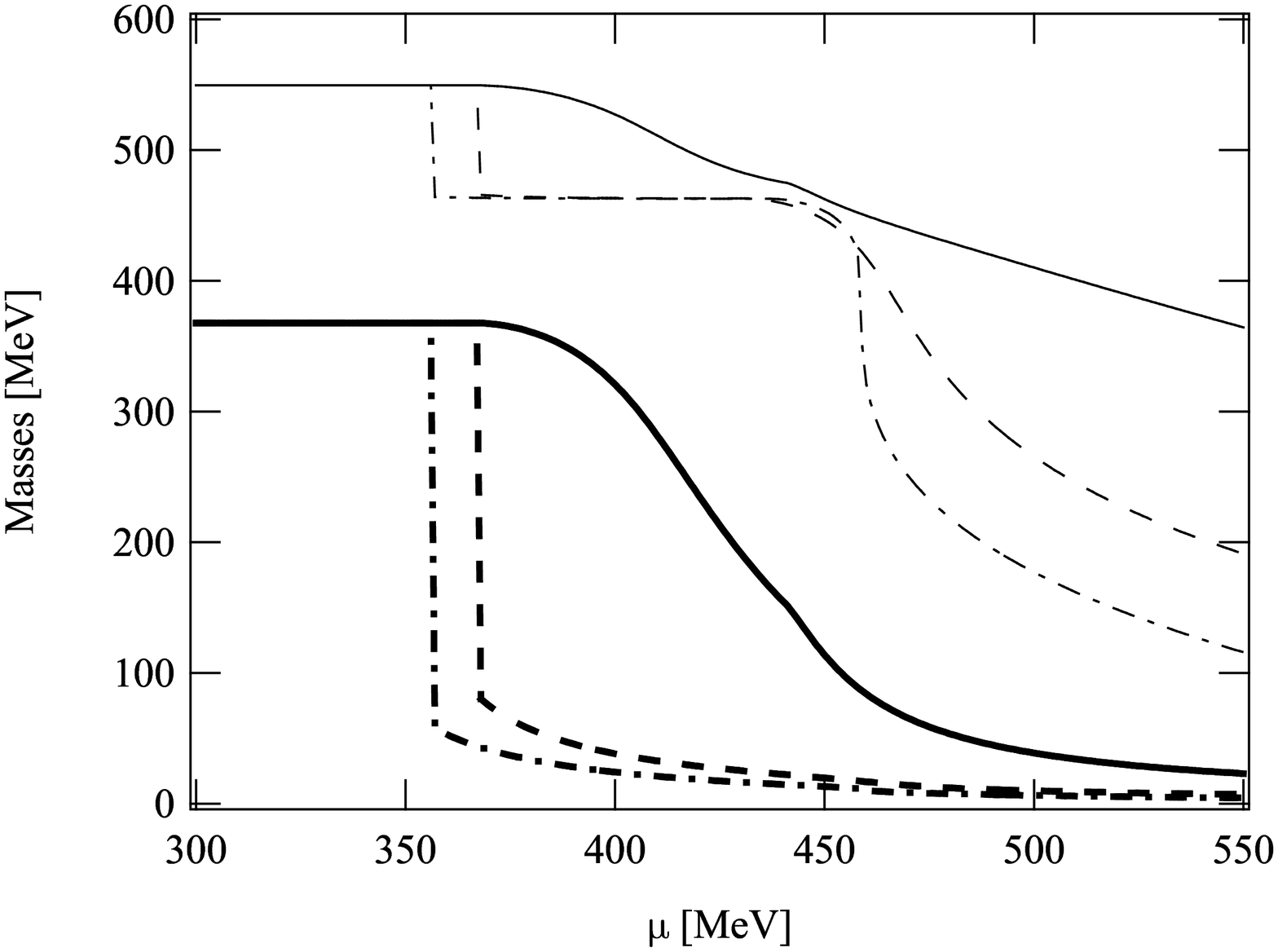}
\includegraphics[width=8cm]{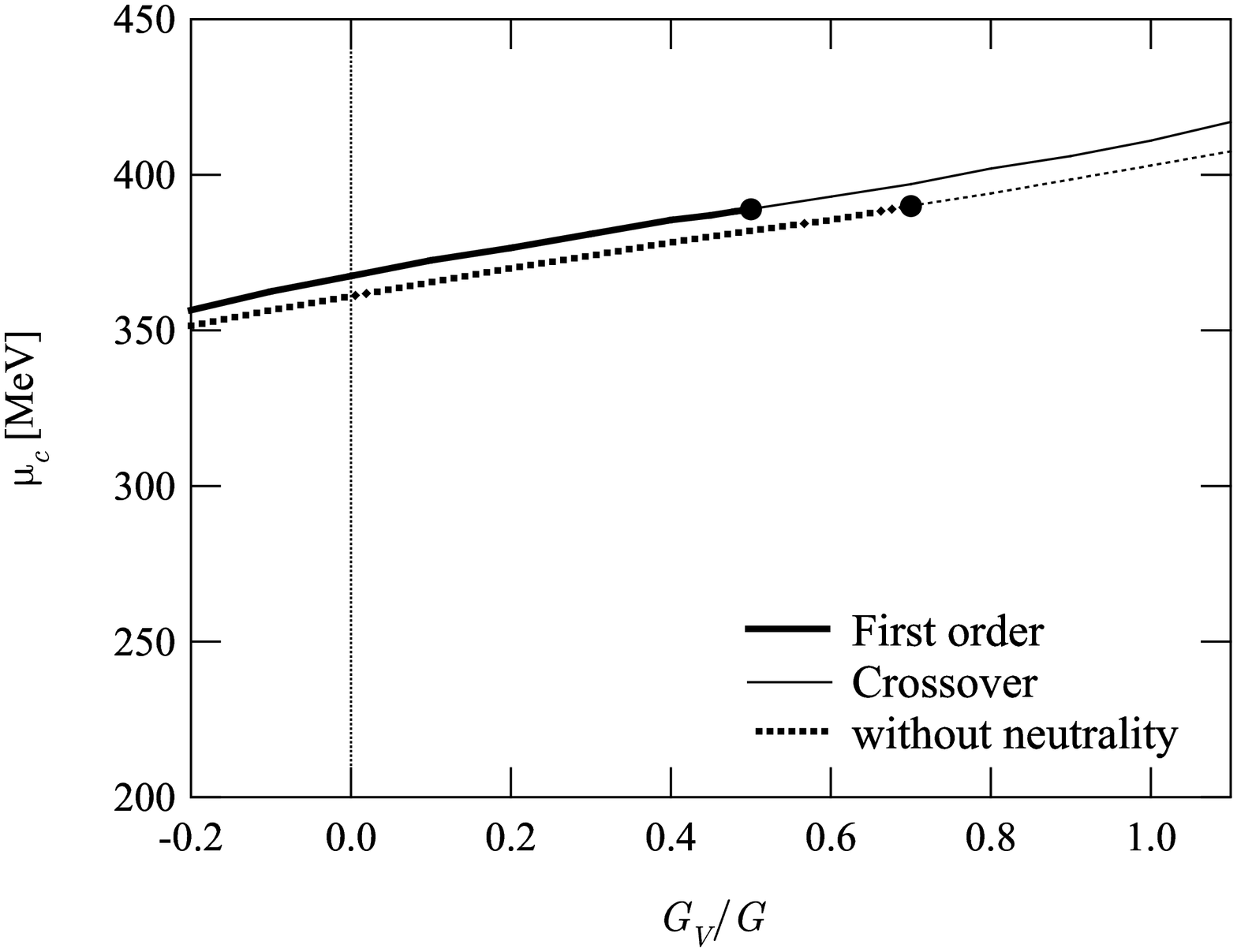}
}
\caption{\label{Fig:MuMu} Left panel: Mean field up quark mass (thick lines) and strange quark mass (thin lines)
against $\mu$ for three different values of the ratio $r_V = G_V/G$ in neutral and $\beta$-equilibrated quark matter:
dot-dashed line corresponds to $r_V = -0.2$, dashed line to $r_V = 0$, solid line to $r_V = +1.1$. Right panel:
critical value of $\mu$ for restoration of the approximate chiral symmetry as a function of $r_V$. Bold line denotes
first order phase transition, solid thin line corresponds to a smooth
 crossover. For comparison, the same quantity computed
 without requiring electrical neutrality is shown by dashed line.}
\end{figure}

The phase diagram of the model can be described equivalently in terms either of the chiral condensates or of the mean
field quark masses. We prefer the latter point of view since it is physically more intuitive. From the qualitative
point of view the behavior of the constituent quark masses as a function of $\mu$ is similar in the cases $r_V = 0$ and
$r_V \neq 0$. In both cases there exist a critical value of $\mu$ such that $M_f$ for $f=$u,d suddenly decreases from
its vacuum value $M_0 \approx 367$ MeV to a lower one, typically a few tens of MeV. However, the order of the
transition depends on the value of $r_V$. For $r_V < 0$ we find a first order transition as in the case $r_V = 0$. In
this case the critical chemical potential is lowered with respect to the case $r_V = 0$. The first order character of
the transition remains even for small and positive values of $r_V$. At $r_V \equiv r_V^c\approx 0.5$ the first order
transition becomes a crossover. In this case we identify the crossover point with the value of $\mu$ where $|d
M_{u}/d\mu|$ is maximum. The magnitude of $|d M_{u}/d\mu|$ at the critical point is lowered as $r_V$ is increased above
the $r_V^c$,
indicating that the crossover becomes more smooth.  These results are in qualitative agreement with those obtained
within the non neutral and non $\beta$-equilibrated VENJL~\cite{Klimt:1990ws,Lutz:1992dv}; they are also in agreement
with the results obtained within the three flavor PNJL model in Ref.~\cite{Fukushima:2008wg}, where it is shown that
the first order line disappears from the $T-\mu$ phase diagram when $G_V$ is increased above a critical value. We
summarize these results in Fig.~\ref{Fig:MuMu} where we plot $M_u$ and $M_s$ against $\mu$ for three representative
values of $r_V$ (left panel), as well as the critical chemical potential as a function of $r_V$ (right panel). For
comparison we plot, by dashed line in the right panel of Fig.~\ref{Fig:MuMu}, the critical chemical potential computed
without requiring the neutrality condition and by setting $\mu_e = 0$. We notice that the neutrality condition has an
effect similar to that of the repulsive vector interaction: it stabilizes the chiral symmetry broken phase with respect
to the restored phase, as it increases the critical chemical potential upwards. The same effect can also be induced by
increasing the strength of repulsive vector interaction. Interestingly enough, the value of critical chemical potential
at which the first order phase transition turns into a smooth crossover is not so much affected by the neutrality
condition. However, the required value of the strength of repulsive vector interaction is significantly reduced if the
condition of neutrality is taken into account. As a consequence, the neutrality constraint actually helps to make the
transition smoother at fixed vector coupling.

\begin{figure}[t!]
\centerline{
\includegraphics[width=10cm]{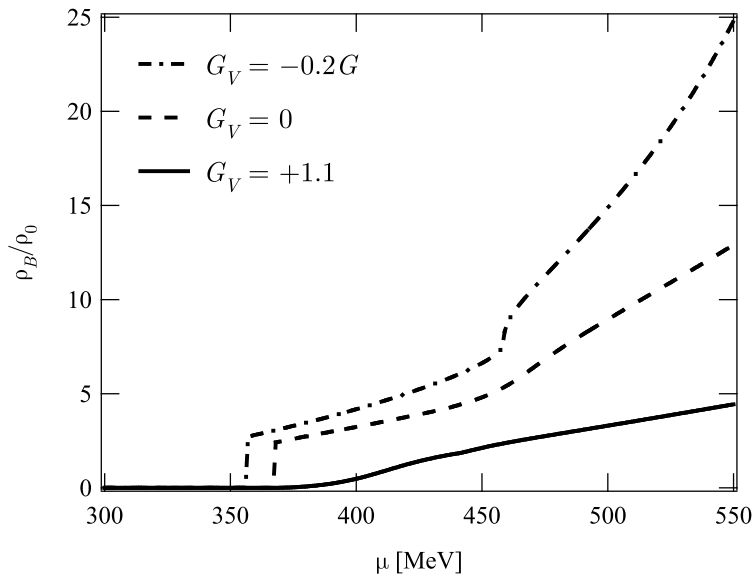}
}
\caption{\label{Fig4} Baryon density in neutral quark matter, in units of the saturation density $\rho_0 = 0.16$
fm$^{-3}$, against $\mu$ for four different values of the ratio  $r_V = G_V/G$: dot-dashed line corresponds to $r_V =
-0.2$, dashed line to $r_V = 0$ and solid line to $r_V = 1.1$. }
\end{figure}

In Fig.~\ref{Fig4} we plot the baryon density $\rho_B$ defined as
\begin{equation}
\rho_B = \frac{1}{3}\left(\rho_u + \rho_d + \rho_s\right)~,
\end{equation}
as a function of $\mu$ for four different values of the ratio $r_V = G_V/G$. The densities of each flavor are computed
by means of Eq.~\eqref{eq:num} with the values of $\mu_e$ and $M_f$ obtained by the minimization procedure. In the
figure the dot-dashed line corresponds to $r_V = -0.2$, dashed line to $r_V = 0$, solid line to $r_V = 1.1$. An
interesting
consequence of the change from first order phase transition 
to crossover into the approximate chiral restored phase when $r_V \ge 0.5$
is that the baryon density as a function of $\mu$ is a continuous
function of $\mu$. Thus the system smoothly passes from a dilute Fermi
gas to a dense one, the size of the smoothness depending on the precise
value of $r_V$. This does not happen when the transition is of first order.

\begin{figure}[t!]
\centerline{
\includegraphics[width=8cm]{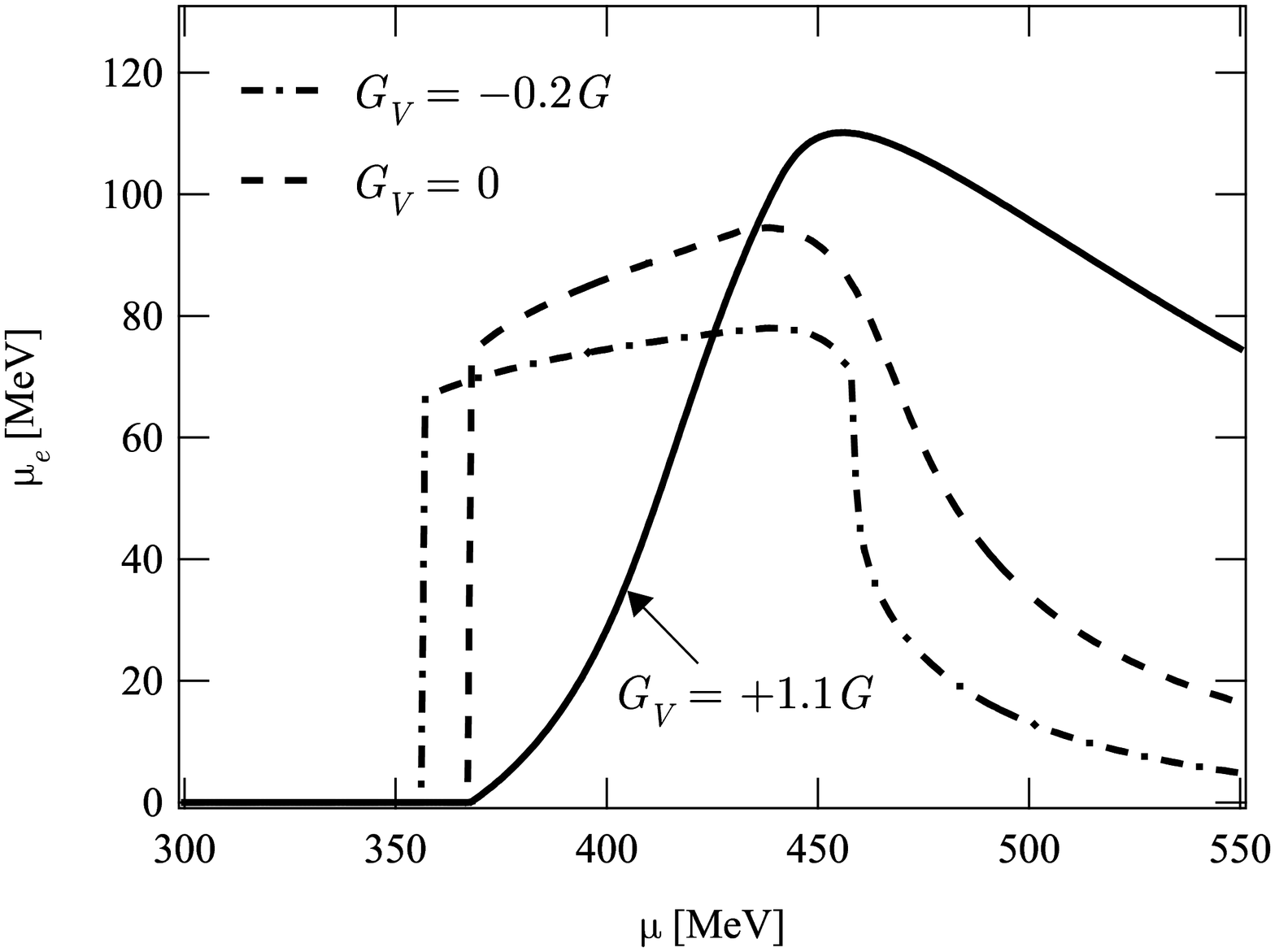}
\includegraphics[width=8cm]{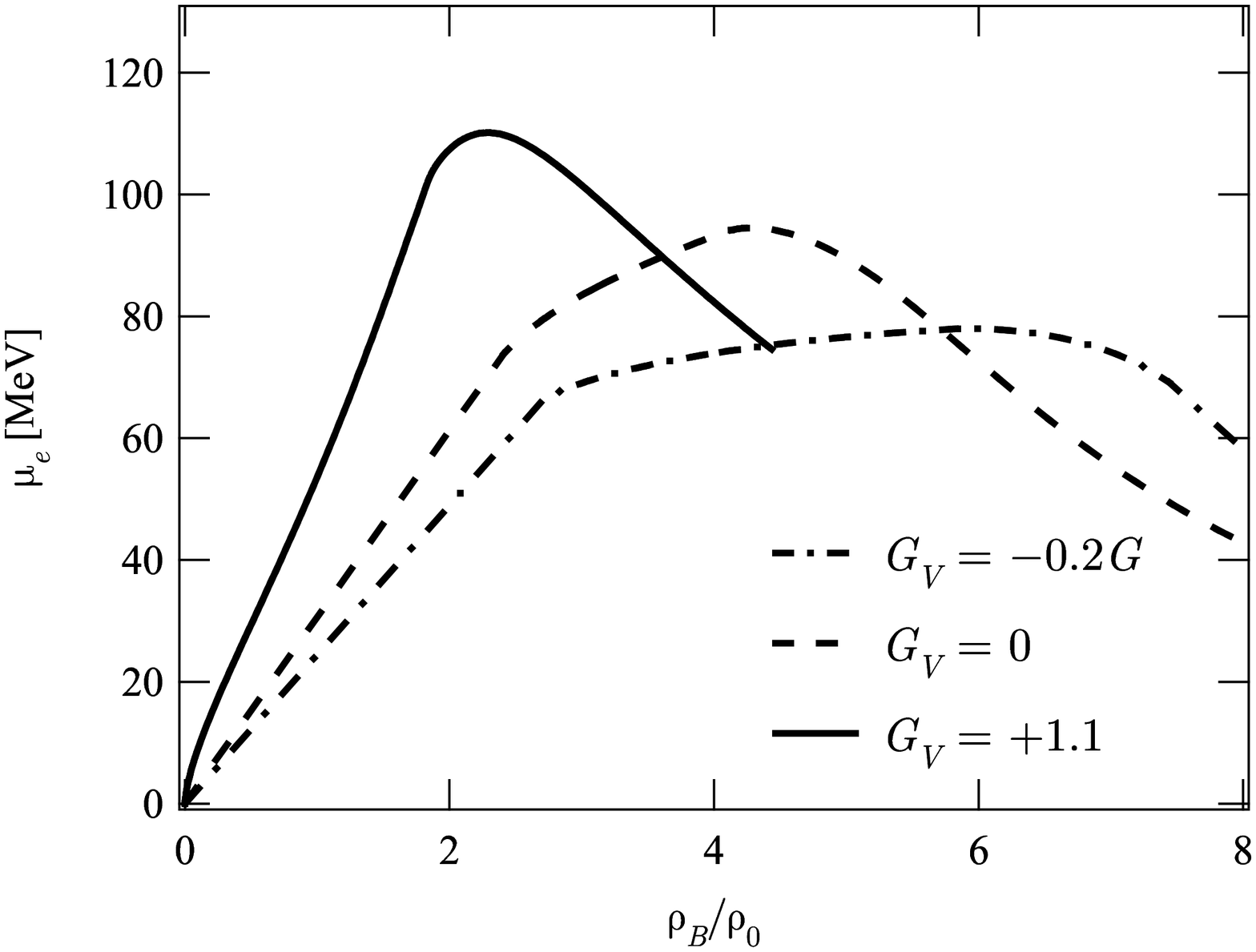}
}
\caption{\label{Fig6} Left panel: electron chemical potential in neutral
 quark matter against $\mu$ for four different values of the ratio $r_V
 = G_V/G$: dot-dashed line corresponds to $r_V = -0.2$, dashed line to
 $r_V = 0$, solid line to $r_V = 1.1$. Right panel: electron chemical
 potential against baryon density. The line sketch is as in the left
 panel. The change of the sign of the slope of the curves at large $\mu$
 (or $\rho_B$) occurs in correspondence of the creation of strange
 quark Fermi spheres ($\mu_s > M_s$).}
\end{figure}

In the left panel of Fig.~\ref{Fig6} we plot the electron chemical potential of neutral quark matter as a function of
$\mu$ for some value of $r_V$. In the right panel we plot the same chemical potential against the baryon density
$\rho_B$. The latter plot is obtained by assembling data from the left panel and from Fig.~\ref{Fig4}. It is
interesting to notice that at a given value of $\mu$, the larger the magnitude of $r_V$ the larger the numerical value
of $\mu_e$. For example at $\mu = 440$ MeV we find $\mu_e(r_V = 0) \approx 95$ MeV, to be compared with $\mu_e(r_V =
+1.1) \approx 120$ MeV. For comparison, at the same value of $\mu$ we find $\mu_e(r_V = -0.2) \approx 80$ MeV. The
change of the sign of the slope of the curves at large $\mu$ (or $\rho_B$) occurs in correspondence of the condition
$\rho_s > 0$ (at lower values of $\mu$ we find $\rho_s = 0$), which implies that strange quarks take a role in the
neutralization of the system and a less number of electrons is needed.

\section{Meson energies and (absence of) condensation}
In this section we compute the pseudoscalar meson energies as a function of the mean quark chemical potential $\mu$ at
$T=0$. To achieve this result we solve the pole equation in the rest frame for the appropriate channel as discussed in
a previous section. We focus on the charged modes here because they are interesting in the context of meson
condensation driven by the electron chemical
potential~\cite{Kogut:2001id,Barducci:2004nc,Warringa:2005jh,Ramos:2000dq,He:2005nk,Abuki:2008wm,Ebert:2006uh,Ebert:2005wr,Ebert:2005cs,Muto:1992pq}.

\subsection{Meson energies and in-medium meson propagator}
Keeping into account the results discussed in the previous section (see
Eq.~\eqref{eq:l}) we write the equation for the energy of the charged kaons as
\begin{equation}
\begin{array}{rcl}
\Re F_{K^\pm}(\omega)&=&0, \\[1ex]
F_{K^\pm}(\omega)&\equiv&1-2K_4^{(+)}\Pi_{K^\pm}(\omega,{\bm 0}),\\[1ex]
\end{array}
\label{eq:poleKm}
\end{equation}
where the polarization function is defined as
\begin{equation}
\Pi_{K^\pm}(\omega,{\bm Q}) = -2i N_c \int \frac{d^4 p}{(2\pi)^4} \text{tr}\left[\gamma_5 \frac{1}{h_u(p)} \gamma_5
\frac{1}{h_s(p + Q)}\right] ~,\label{eq:64}
\end{equation}
where $Q=(\omega,{\bm Q})$ with ${\bm Q}$ being the three momentum of Kaon,
and $\Re$ denotes the real part. In the above equation the trace is
understood on Dirac indices only. Performing the
trace the pole equation~\eqref{eq:poleKm} in the rest frame reads:
\begin{equation}
24 \Re \int \frac{d^3 p}{(2\pi)^3}~T\!\sum_{n=-\infty}^{+\infty} \frac{M_u M_s + {\bm p}^2  - (i\omega_n +
\mu_u)(i\omega_n + i\Omega_m + \mu_s)}{\left[(i\omega_n + \mu_u)^2 - {\bm p}^2 - M_u^2\right]\left[(i\omega_n +
i\Omega_m + \mu_s)^2 - {\bm p}^2 - M_s^2\right]} = \frac{1}{2K_4^{(+)}}~. \label{eq:poleKm2}
\end{equation}
Once again we have introduced a finite temperature $T$ in order to handle infrared divergencies that arise when the
chemical potential of the flavor $f$ equals its mean field mass $M_f$. At the end of the calculation we put
$T\rightarrow 0^+$. In the above equation $\Omega_m = \pi T n$ is the boson Matsubara frequency.

The retarded real time propagator is obtained via $i\Omega_m \rightarrow \omega +i 0^+$ after summation over fermion
Matsubara frequencies. The result turns out to depend on the external energy $\omega$ only on the combination
\begin{equation}
Q_0 = \omega + \mu_s - \mu_u = \omega + \mu_e +4G_V(\rho_u - \rho_s)~, \label{eq:Qtilde}
\end{equation}
Thus also the kaon lagrangian in momentum space depends only on $Q_0$. This implies that in the derivative expansion
one can build only terms that contain the covariant derivative
\begin{equation}
iD_0 \equiv i\partial_0 + \mu_e +4G_V(\rho_u - \rho_s)~.\label{eq:ppp}
\end{equation}

We make analytically the sum over Matsubara frequencies in Eq.~\eqref{eq:poleKm2}. After continuation to real external
energies we take the limit $T\rightarrow 0^+$, and we get an expression that depends on energy only through
$Q_0$. We do analytically the integral over $|\bm p|$. Finally we solve Eq.~\eqref{eq:poleKm2} in the variable
$Q_0$. The solution of the pole equation in the $Q_0$ variable defines the in-medium kaon mass
$m^\star_{K^{\pm}}$. Within our convention in Eq.~\eqref{eq:64} the
positive (respectively negative) energy solution of
Eq.~\eqref{eq:poleKm2}, $Q_0=m_{K^-}^\star$
(respectively $Q_0=-m_{K^+}^\star$) correspond to the $K^-$
(respectively $K^+$) in-medium mass.
Once the pole equation is solved at a given value of $\mu$, we obtain
the in-medium energies trivially from Eq.~\eqref{eq:Qtilde}:
\begin{eqnarray}
\omega_{K^{+}} &=& \mu_e + 4G_V(\rho_u - \rho_s) + m_{K^+}^\star, \label{eq:i1}\\
\omega_{K^{-}} &=& -\mu_e - 4G_V(\rho_u - \rho_s) + m_{K^-}^\star~. \label{eq:i2}
\end{eqnarray}

At $G_V = 0$ the chemical potential felt by kaons is $-q \mu_e$ where $q$ denotes the electric charge (in units of the
electron charge) of the meson. When $G_V$ is switched on, the kaon chemical potential is shifted from $-q \mu_e$ by
virtue of the strong interactions. As a matter of fact, since the quark densities are related to the expectation values
of the $\rho$, $\omega$ and $\phi$ mesons, see Eqs.~\eqref{eq:rhoU} and~\eqref{eq:rhoS}, the $G_V$ terms in the meson
energies can be interpreted as due to the interactions of kaons with the aforementioned vector mesons (there is also a
modification of $m_K^\star$ when $G_V$ is switched on). We stress however that these interactions do not exist at the
tree level within the model we study in this work, but arise only as loop effects: the quarks interact with the
expectation values of the vectors and the quark loops, that generate the mass and the kinetic terms of the pseudoscalar
mesons, will depend on these expectation values. Another source of interactions among pseudoscalar and vectors (as well
as axial-vectors) are the terms of the cubic order in meson fields in the loop expansion of Eq.~\eqref{eq:APP1}, that
we do not consider in this study for simplicity and will be the subject of a next paper.

The other pseudoscalar channels are treated in a similar way. The pole equation for the charged pions reads
\begin{equation}
24 \Re\int \frac{d^3 p}{(2\pi)^3}~T\!\sum_{n=-\infty}^{+\infty}
 \frac{M_u M_d + {\bm p}^2  - (i\omega_n+\mu_u)(i\omega_n + i\Omega_m +
 \mu_d )}{\left[(i\omega_n + \mu_u)^2 - {\bm p}^2 -
       M_u^2\right]\left[(i\omega_n + i\Omega_m + \mu_d)^2 - {\bm
       p}^2 - M_d^2\right]} = \frac{1}{2K_1^{(+)}}~,
       \label{eq:polePm2}
\end{equation}
and in this case the polarization tensor depends on external energy only through the combination
\begin{equation}
Q_0 = \omega + \mu_d - \mu_u = \omega + \mu_e +4G_V(\rho_u - \rho_d)~.
\end{equation}
In this way we have
\begin{eqnarray}
\omega_{\pi^{+}} &=& \mu_e + 4G_V(\rho_u - \rho_d) + m_{\pi^+}^\star~, \label{eq:i3}\\
\omega_{\pi^{-}} &=& -\mu_e - 4G_V(\rho_u - \rho_d) + m_{\pi^-}^\star~. \label{eq:i4}
\end{eqnarray}
Similarly, the equation for the neutral kaons is given by
\begin{equation}
24 \Re \int \frac{d^3 p}{(2\pi)^3}~T\!\sum_{n=-\infty}^{+\infty} \frac{M_d M_s + {\bm p}^2  - (i\omega_n +\mu_d
)(i\omega_n + i\Omega_m + \mu_s)}{\left[(i\omega_n + \mu_d)^2 - {\bm p}^2 - M_d^2\right]\left[(i\omega_n + i\Omega_m +
\mu_s)^2 - {\bm p}^2 - M_s^2\right]} = \frac{1}{2K_6^{(+)}}~, \label{eq:poleK02}
\end{equation}
with
\begin{equation}
Q = \omega + \mu_s - \mu_d = \omega + 4G_V(\rho_d - \rho_s)~.
\end{equation}
The in-medium energies are given by
\begin{eqnarray}
\omega_{K^{0}} &=& 4G_V(\rho_d - \rho_s) + m_{K^0}^\star~, \label{eq:i5}\\
\omega_{\bar{K^{0}}} &=&  - 4G_V(\rho_d - \rho_s) + m_{\bar{K}^0}^\star~. \label{eq:i6}
\end{eqnarray}

\begin{figure}[t!]
\begin{center}
\includegraphics[width=8cm]{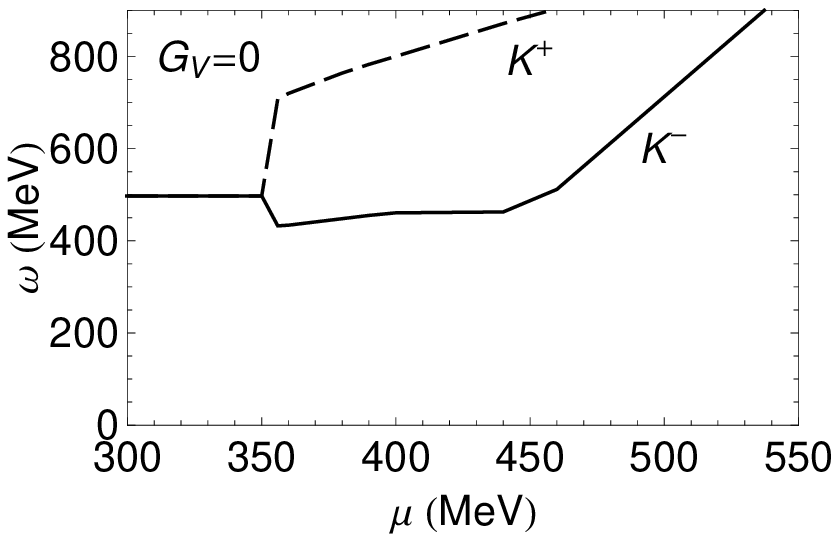}~~~~~\includegraphics[width=8cm]{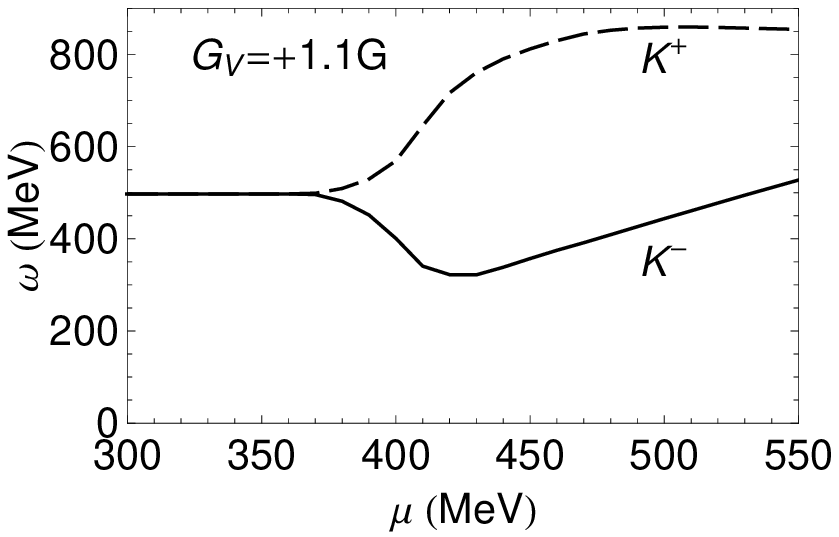}\\
\includegraphics[width=8cm]{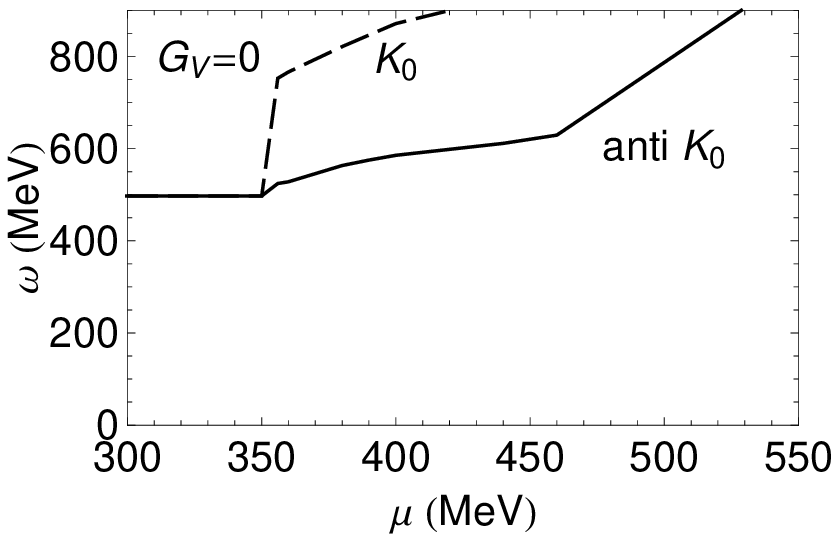}~~~~~\includegraphics[width=8cm]{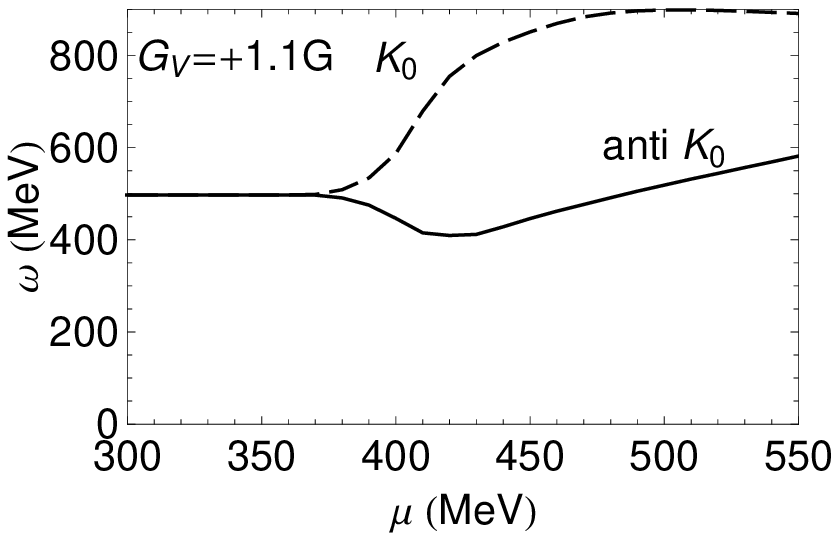}\\
\includegraphics[width=8cm]{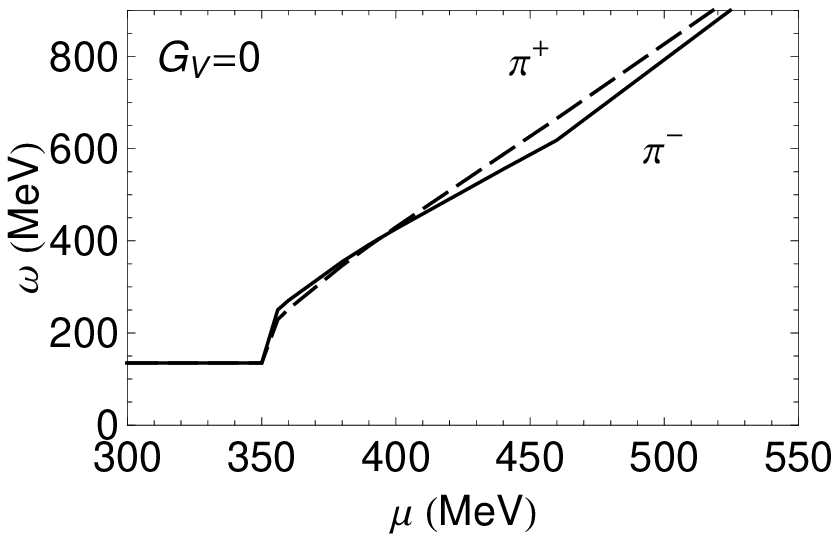}~~~~~\includegraphics[width=8cm]{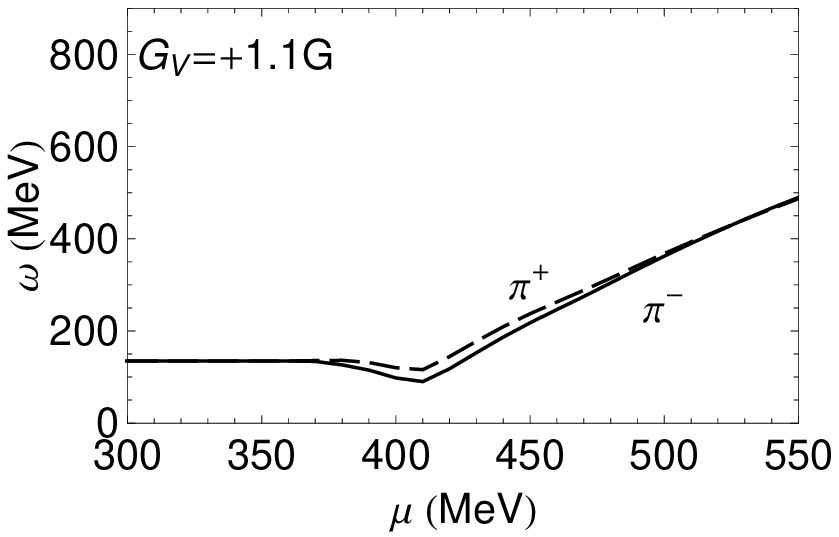}
\end{center}
\caption{\label{Fig8} Solutions of the pole equations in neutral quark matter for charged kaons (upper panel), neutral
kaons (middle panel) and charged pions (lower panel) as a function of $\mu$ for two different values of $r_V$. }
\end{figure}

\begin{figure}[t!]
\centerline{
\includegraphics[width=0.4\textwidth]{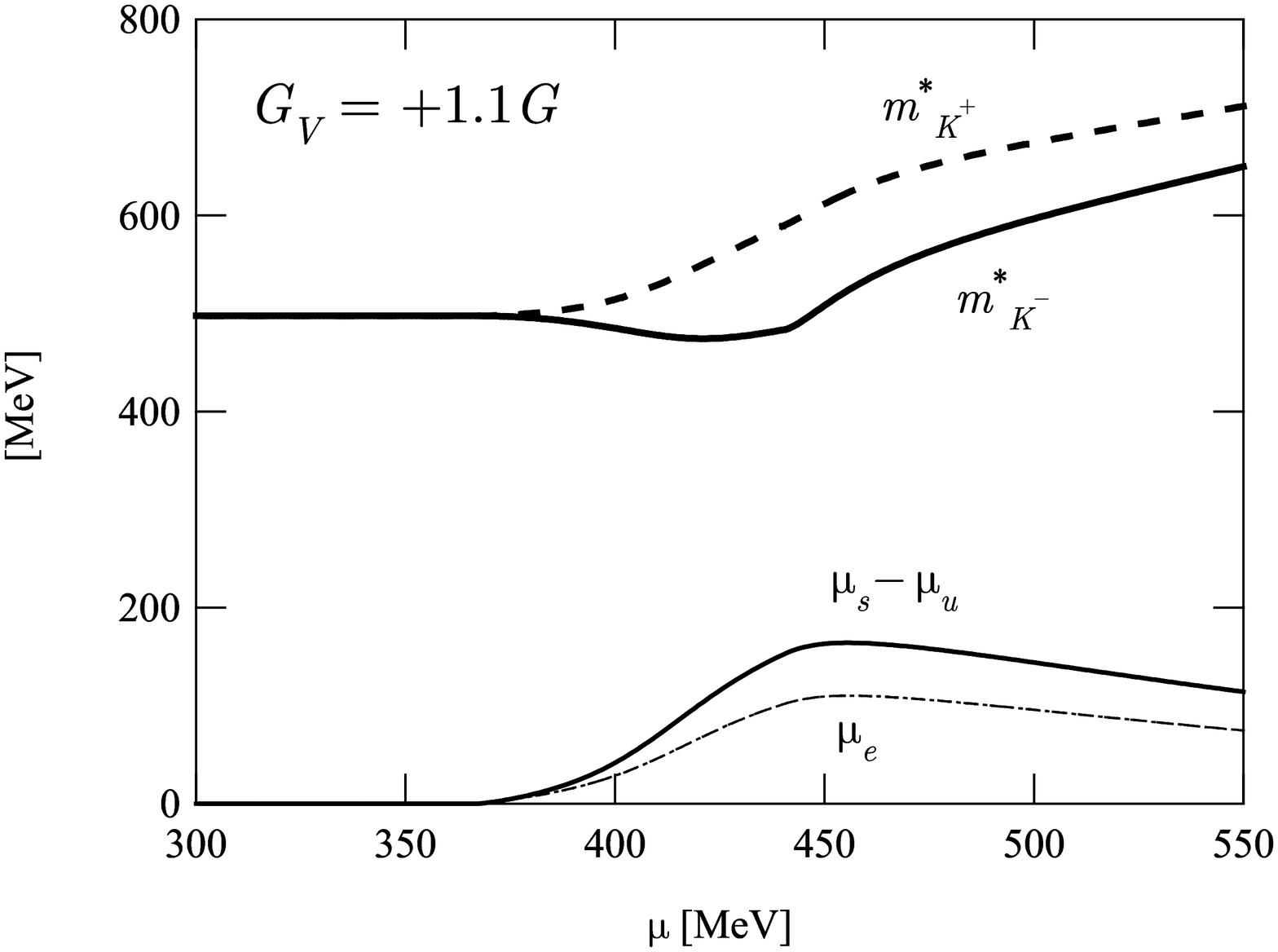}
} \caption{The various contributions to in-medium Kaon energies. Effective in-medium Kaon masses $m_{K_\pm^\star}$, the
electric chemical potential $\mu_e$ and $\mu_s-\mu_u$ as a function of $\mu$. $m_{K_\pm\star}\pm(\mu_s-\mu_u)$
correspond to the two curves in the top-right figure of Fig.~\ref{Fig8}. The deviation of $\mu_s-\mu_u$ from $\mu_e$
comes from the repulsive vector interaction. } \label{fig:each}
\end{figure}

We solve Eqs.~\eqref{eq:poleKm2},~\eqref{eq:polePm2} and~\eqref{eq:poleK02} by using the values of $\sigma_f$ and
$\mu_e$ that correspond to the neutral global minimum of the effective
potential~\eqref{eq:totOmega}. In Fig.~\ref{Fig8} we plot the solutions
of the real part of the pole equations for charged kaons (upper panels),
corresponding to Eqs.~\eqref{eq:i1},~\eqref{eq:i2}; neutral kaons
(middle panels), corresponding to Eqs.~\eqref{eq:i3},~\eqref{eq:i4}; and
charged pions (lower panels), corresponding to
Eqs.~\eqref{eq:i5},~\eqref{eq:i6}. They are plotted against $\mu$ for
$r_V = 0$ (left panels) and $r_V = +1.1$ (right panels). Results for
intermediate values of $r_V$ do not differ qualitatively from those
shown in the figure. Also, for negative values of $r_V$ the difference
with the case $r_V= 0$ is that the location of the 
discontinuity in the in-medium
energies is shifted to lower values of the quark chemical potential.

The plots of Fig.~\ref{Fig8} are interesting for several reasons. A
general feature of the vector interaction is the stabilization of the
charged pseudoscalar states: the comparison at the same chemical
potential of the in-medium energies of a meson with and without keeping
into account the repulsive vector interaction shows that they are higher
in the latter case than in the former one. Secondly, we notice that the
effect of the vector interaction on the $K^-$
and $\bar{K^0}$ in-medium energies is to lower them in the $\mu$ regime
where chiral symmetry is approximately
restored. For example at $\mu=460$ MeV we find $\omega_{K^-} \approx
406$ MeV for $r_V = 0$, to be compared with $\omega_{K^-} \approx 290$
MeV for $r_V = 1$. This is in part due to a mild lowering of $m^\star_K$
as $G_V$ is increased, but mainly to the effective chemical potential
arising from interaction of $K^-$ and $\bar{K^0}$ with the expectation
values of the vector meson fields via the quasi-quarks in the loop, as
is clear in the expressions, Eq.~(\ref{eq:i2}) and Eq.~(\ref{eq:i6}).
In the case of $K^-$ there is still a further lowering of the energy due to the larger value of $\mu_e$,
see Fig.~\ref{Fig6}: at $\mu = 460$ MeV we find $\mu_e \approx 95$ MeV for $r_V = 0$, to be compared with $\mu_e
\approx 120$ MeV for $r_V = +1.1$.

For completeness, in Fig.~\ref{fig:each} we plot the separate contributions to the in-medium Kaon energies. The solid
lines indicated by $m_{K_\pm^\star}$ express the in-medium Kaon masses while $\mu_e$ is electron chemical potential.
The in-medium kaon mass $m_{K_-^\star}$ should be compared with $\mu_s-\mu_u=\mu_e+4G_V(\rho_u-\rho_s)$, which is
effective chemical potential felt by $K^-$: it is shifted from the bare charge chemical potential $\mu_e$ means of the
vector interaction. Both the electron chemical potential and the vector interaction tend to lower the in-medium $K^{-}$
energy, even though the lowering is not enough for the kaon condensate to form.
\begin{figure}[t!]
\centerline{
\includegraphics[scale=0.38]{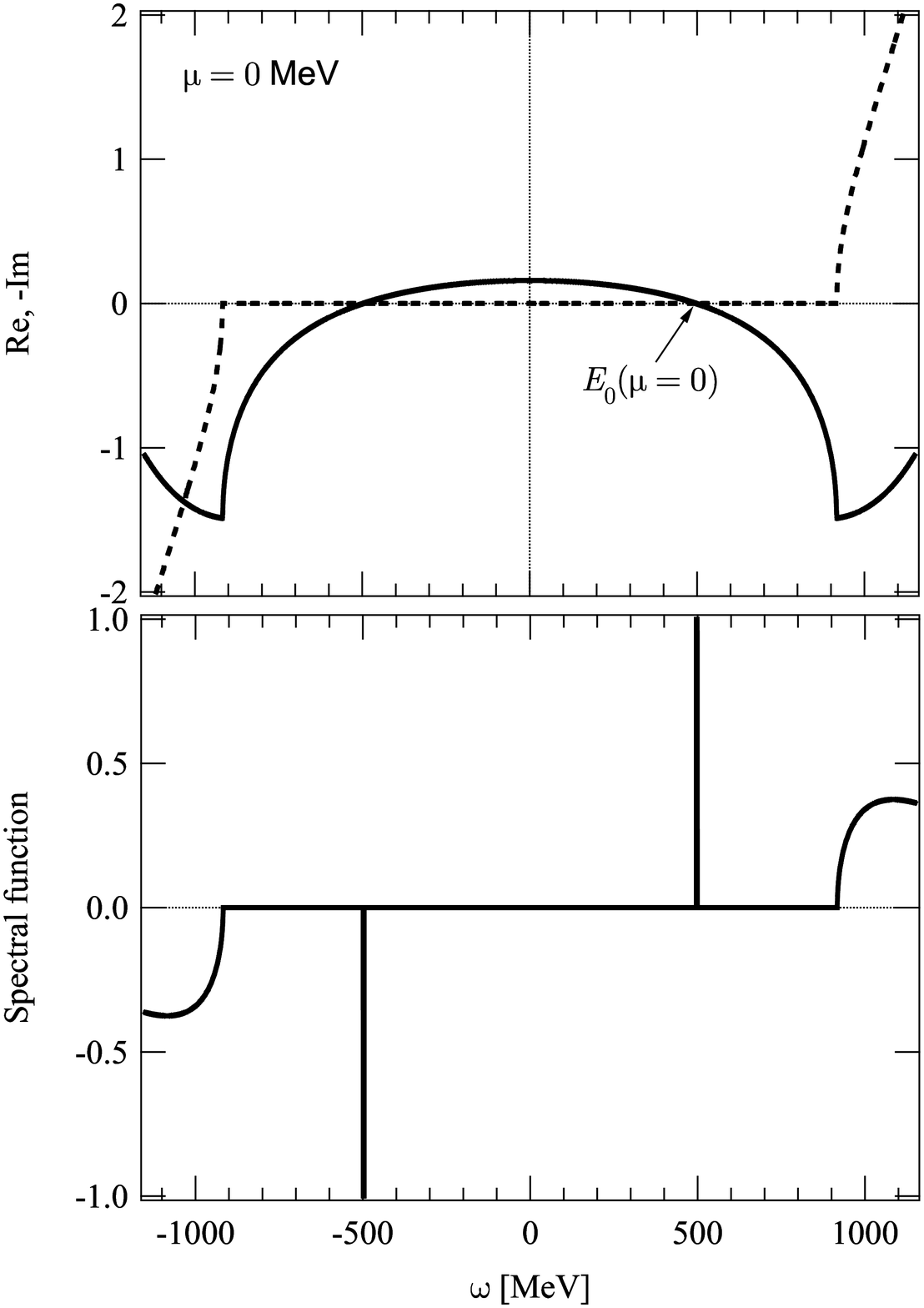}
\includegraphics[scale=0.38]{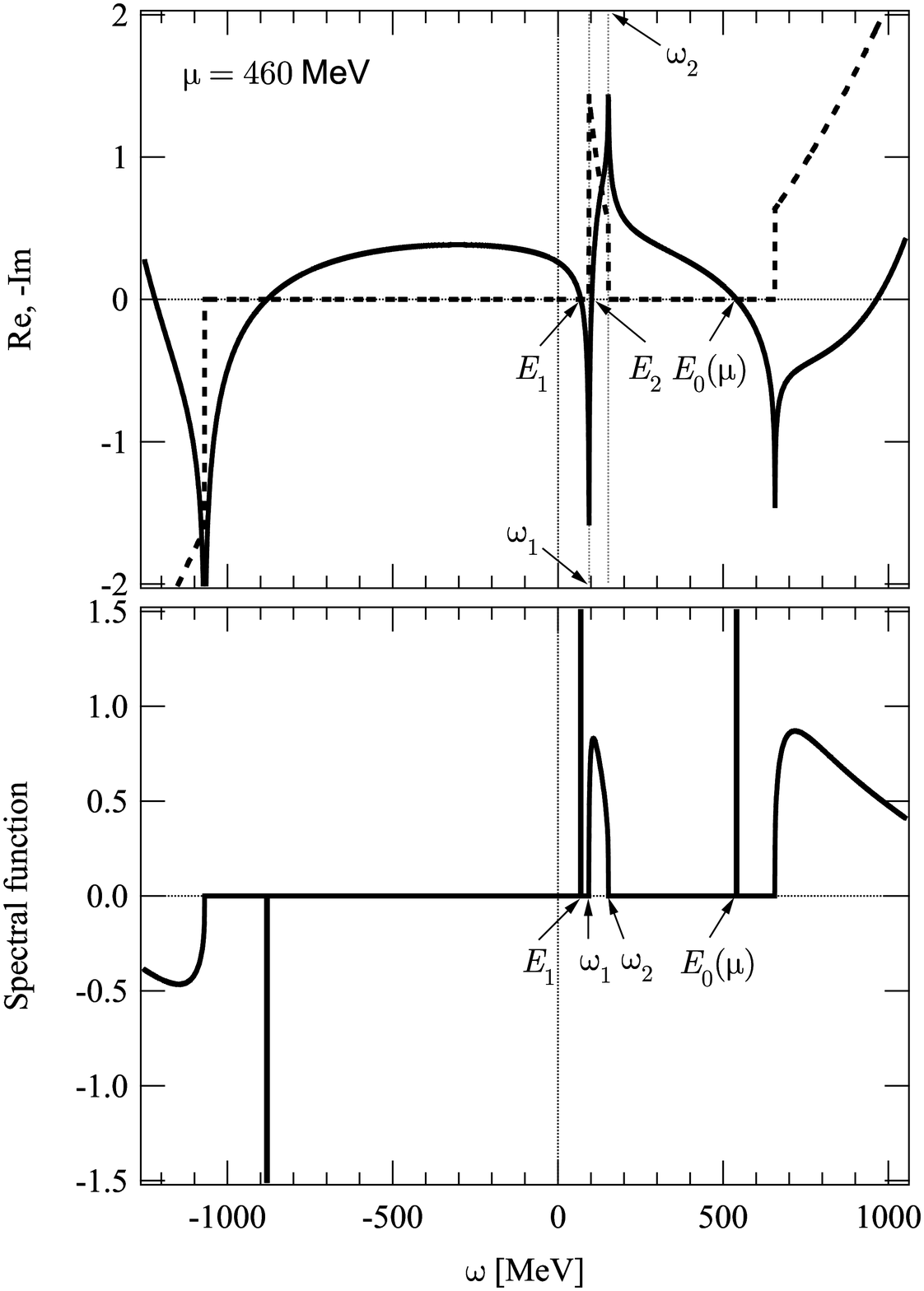}
}
\caption{Upper panel:~Real and (muinus) imaginary part of
 $F_{K^\pm}(\omega)\equiv1-2K_4^{+}\Pi_{K^\pm}(\omega,\bm q =0)$ as a function of
$\omega$ for $G_V = +1.1 G$, for two different values of $\mu$. Solid
 line corresponds to the real part, dashed line to
the imaginary part. Lower panel:~Corresponding spectral function
of Kaonic propagator $\rho(\omega)=\Im\left({1}/{F_{K^\pm}(\omega)}\right)$.}
\label{Fig:KmProp}
\end{figure}

\begin{figure}[t!]
\centerline{
\includegraphics[width=0.4\textwidth]{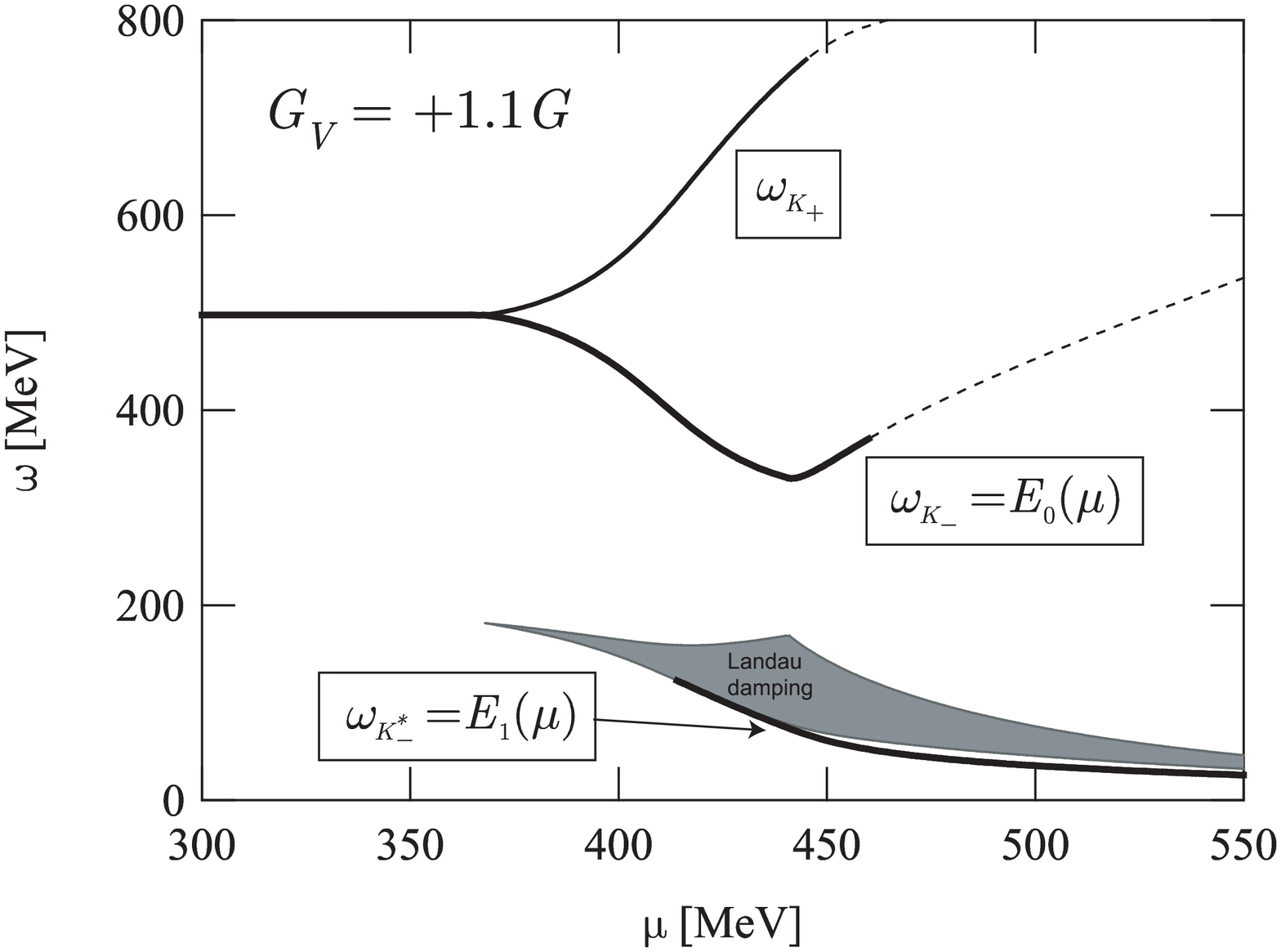}
}
\caption{The energy of the collective modes in the ($\omega$-$\mu$) plane.
The new collective mode with quantum number $K^-$ induced by
the Landau damping. The shaded area corresponds to the region where the
responce function suffers from imaginary part due to the Landau damping.
$\omega_{K^-}$ and $\omega_{K^+}$ are the same as two curves in the
 top-right figure of Fig.~\ref{Fig8}. The points where the solid lines
 turn into the dashed ones indicate the thresholds for the continuum:
 across the points, the modes become no longer stable decaying into the
 constituents.}
\label{fig:spectral}
\end{figure}

The shape of the inverse $K^-$ propagator at rest as a function of energy and $\mu$ has an interesting feature. In
Fig.~\ref{Fig:KmProp} we plot the real and the imaginary part of
$F_{K^\pm}(\omega)\equiv1-2K_4^{+}\Pi_{K^\pm}(\omega,\bm q =0)$ for $r_V = +1.1$ and two values of $\mu$. They are
represented respectively by the solid and the dashed line. At $\mu = 0$ we have $\Re F_{K^\pm}(m_K)=\Re
F_{K^\pm}(-m_K)=0$ and $\Im F_{K^\pm}(\omega) = 0$ for $|\omega| < M_u +M_s$. Here $m_K$ denotes the vacuum kaon mass.
As $\mu$ is increased above the chiral transition then $F_{K^\pm}(\omega)$ develops two singularities at intermediate
values of $\omega$, see the right panel of Fig.~\ref{Fig:KmProp}.  Moreover two new zeros appear. Thus in principle at
large $\mu$ four kaon modes appear. The solutions we have chosen to draw Fig.~\ref{Fig:KmProp} are those with the
larger magnitude.

In order to make clear the physical content of the intermediate singularities at $\omega=\omega_1, \omega_2$, we will
take a detailed look at the imaginary part of $F_{K^\pm}(\omega)$ from now on. It is possible to find some analytic
formula for the imaginary part which is physically related to the $K^-$ decay rate in the medium. By use of the formula
$\Im(1/(\omega+x+i\delta))=-i\pi\delta(\omega+x)$, we find
\begin{equation}
\begin{array}{rclcc}
  \Im F_{K^\pm}(\omega)&=&\displaystyle-\frac{6K_4^{(+)}}{\pi}%
   \int_0^\infty d p\frac{E_uE_s+M_uM_s+p^2}{E_u
   E_s}(1-f_{\bar{u}}-f_s)\delta(Q_0-E_u-E_s)&\cdots\cdots& (a)\\[1ex]
  & &\displaystyle+\frac{6K_4^{(+)}}{\pi}\int_0^\infty
   d p\frac{-E_uE_s+M_uM_s+p^2}{E_u E_s}(f_{u}-f_{s})%
   \delta(Q_0+E_u-E_s)&\cdots\cdots&(b)\\[1ex]
  & &\displaystyle+\frac{6K_4^{(+)}}{\pi}\int_0^\infty
   d p\frac{-E_uE_s+M_uM_s+p^2}{E_u E_s}%
   (f_{\bar{s}}-f_{\bar{u}})\delta(Q_0-E_u+E_s)%
  &\cdots\cdots&(c)\\[1ex]
  & &\displaystyle+\frac{6K_4^{(+)}}{\pi}%
     \int_0^\infty d p\frac{E_uE_s+M_uM_s+p^2}%
     {E_uE_s}(1-f_{u}-f_{\bar{s}})\delta(Q_0+E_u+E_s),&\cdots\cdots&(d)\\[1ex]
\end{array}
\end{equation}
where $Q_0$ is again the shifted energy defined by Eq.~(\ref{eq:Qtilde}), $E_u=\sqrt{M_u^2+p^2}$ and
$E_s=\sqrt{M_s^2+p^2}$ are the u and s quark energies, and $f'$s are the Fermi blocking factors defined by
$f_{i}=\theta(\mu_i- E_i)$ for the quark fermi distribution, and $f_{\bar{i}}=\theta(-\mu_i - E_i)$ for the antiquark
Fermi distribution. Since the momentum integral is definitely convergent, we have removed the cutoff from the
evaluation of the imaginary part. As easily guessed from the blocking factors and delta function guaranteeing the
energy conservation, $(a)$ expresses the two body decay from $K^-$ to $(s\bar{u})$, $(b)$ does the Landau damping
process of $K^-$ absorbing onshell u quark decaying into s quark, $(c)$ represents the inverse Landau damping of $K^-$
absorbing anti s quark decaying into anti u quark, and $(d)$ is for the decay of $K^-$ by absorbing u quark and anti s
quark. The $p$-integral is extremely trivial due to delta functions, yielding
\begin{equation}
\begin{array}{rclcc}
  \Im F_{K^\pm}(\omega)&=&\displaystyle-%
   \bigg[\frac{6K_4^{(+)}p_*}{\pi}%
   \frac{E_u^*E_s^*+M_uM_s+p_*^2}{E_u^*+
   E_s^*}(1-f_{\bar{u}}-f_s)\bigg]%
   \theta\left[Q_0>M_u+M_s\right]%
   &\cdots\cdots& (a)\\[1ex]
  & &\displaystyle+\bigg[\frac{6K_4^{(+)}p_\star}{\pi}%
   \frac{-E_u^\star E_s^\star+M_uM_s+p_\star^2}%
   {|E_u^\star-E_s^\star|}(f_{u}-f_{s})\bigg]%
   \theta\left[0<Q_0<M_s-M_u\right]&\cdots\cdots&(b)\\[1ex]
  & &\displaystyle+\bigg[\frac{6K_4^{(+)}p_\star}{\pi}%
   \frac{-E_u^\star E_s^\star+M_uM_s+p_\star^2}%
   {|E_u^\star-E_s^\star|}%
   (f_{\bar{s}}-f_{\bar{u}})\bigg]\theta\left[-M_s+M_u<Q_0<0\right]%
  &\cdots\cdots&(c)\\[1ex]
  & &\displaystyle+\bigg[\frac{6K_4^{(+)}p_*}{\pi}%
     \frac{E_u^*E_s^*+M_uM_s+p_*^2}%
     {E_u^*+E_s^*}(1-f_{u}-f_{\bar{s}})\bigg]%
     \theta\left[Q_0<-M_u-M_s\right],%
   &\cdots\cdots&(d)\\[1ex]
\end{array}
\label{eq:ImF}
\end{equation}
where $E_i^{*(\star)}\equiv\sqrt{M_i^2+p_{*(\star)}^2}$,
and $\{p_*,p_{\star}\}$ are now to be determined through
the condition of energy conservation,
\begin{equation}
\begin{array}{rclccc}
   E_u^*+E_s^*&=&|Q_0|&\quad&\mbox{for}& |Q_0|>M_u+M_s,\\[1ex]
   E_s^\star-E_u^\star&=&|Q_0|&\quad&\mbox{for}& |Q_0|<M_s-M_u,
\end{array}
\end{equation}
each of which has a unique solution at fixed $Q_0$ in its domain
of definition. It turns out both have the same functional dependence on
$Q_0$, {i.e.,}
\begin{equation}
\begin{array}{c}
   p_{*(\star)}=\displaystyle %
   \frac{\sqrt{((M_s-M_u)^2-Q_0^2)((M_s+M_u)^2-Q_0^2)}}{2|Q_0|}.
\end{array}
\label{eq:pstar}
\end{equation}
We now consider in which condition the Landau damping process labeled by
(b): $u+K^-\to s$, (and the inverse production process $(s\to u+K^-)$)
is possible. As demonstrated above, the energy conservation
$E_s^\star=Q_0+E_u^\star$ has a unique solution when $Q_0$ is in the
interval:
\begin{equation}
  0<Q_0<M_s-M_u.
\label{eq:kinc}
\end{equation}
From Eq.~(\ref{eq:pstar}), we see $p_\star$ is a decreasing function of $Q_0$, and when $Q_0$ approaches zero $p_\star$
diverges. In addition to the energy conservation, the Pauli principle should be also satisfied: when
$p_\star>\sqrt{\mu_u^2-M_u^2}$, or equivallently $E_u^\star>\mu_u$, there is no $u$ quark avaiable to contribute to the
absorption process $u+K^-\to s$. This can be also seen in the blocking factor $f_u-f_s$ in (b) of Eq.~(\ref{eq:ImF}):
This factor comes from the sum of two blocking factors, $f_u(1-f_s)$ for the decay, and $-f_s(1-f_u)$ for the
corresponding creation process, $s\to u+K^-$. Then in order for the total decay rate to be nonzero, either
$\{f_u=1,f_s=0\}$ or $\{f_u=0,f_s=1\}$ should hold. In our case $M_s>M_u$ is always realized so we need to consider
only the former condition, that is, $p_{Fs}<p_\star<p_{Fu}$, where the Fermi momentum is defined as
$p_{Fi}=\sqrt{\mathrm{ max.}(0,\mu_i^2-M_i^2)}$. Solving this condition together under the kinematic constraint
Eq.~(\ref{eq:kinc}), we obtain the condition:
$\sqrt{M_s^2+p_{Fu}^2}-\sqrt{M_u^2+p_{Fu}^2}<Q_0%
<\sqrt{M_s^2+p_{Fs}^2}-\sqrt{M_u^2+p_{Fs}^2}$,
which, combined with $Q_0=\omega+\mu_s-\mu_u$, translates into
\begin{equation}
\begin{array}{rcl}
 \omega_1&=&\sqrt{M_s^2+p_{Fu}^2}-\mu_s-\mathrm{max.}(M_u-\mu_u,0),\\[1ex]
 \omega_2&=&-\sqrt{M_u^2+p_{Fs}^2}+\mu_u+\mathrm{max.}(M_s-\mu_s,0).\\[1ex]
\end{array}
\end{equation}

We have checked that these equations reasonably reproduce the numerical values of $\{\omega_1,\omega_2\}$ in the right
panel of Fig.~\ref{Fig:KmProp}. We conclude that the imaginary part in the range $\omega_1<\omega<\omega_2$ is
non-vanishing due to the Landau damping of $K^-$ by absorption of on-shell u quark in the system. What is more
intriguing is that below this Landau damping threshold, there appears a new pole implying a new collective mode with
the same quantum number $K^-$ as shown in Fig.~\ref{Fig:KmProp}. We have checked that this pole appears only at high
density, $\mu\alt 420$ MeV, as depicted in Fig.~\ref{fig:spectral}. At the moment we cannot discern if this new light
mode is an artifact of the NJL model itself, as well as of the RPA approximation. However, if this is not the case, it
is of great interest since it might change the low temperature behavior of the thermodynamics (e.g., strangeness
population) in a compact star. We leave the detailed investigation of the nature of this new pole to a forthcoming
paper.

A third point that deserves discussion is the missed symmetry in the $K^0/\bar{K^0}$ spectrum. This splitting is
present even in the case $G_V = 0$. It is due in this case to the fact that $K^0\sim (d\bar{s})$ and $\bar{K}_0 \sim
(s\bar{d})$: onshell d quarks in the system tend to reduce the binding energy for $K_0$ state by Pauli blocking. This
is the reason why we see that $M_{K_0}$ is higher than $K_{\bar{K}_0}$. The splitting is more pronounced by the vector
interaction, see right panel in Fig.~\ref{Fig8} because of the induced effective
chemical potential, 
Eqs.~\eqref{eq:i5} and~\eqref{eq:i6}.

We finally comment on the absence of kaon condensation in our model. It
is well established that at in the range of baryon densities $(2.3-5)\rho_0$
and for neutral and $\beta$-equilibrated nuclear matter,
kaon condensation occurs, see
for example~\cite{Thorsson:1993bu,Brown:2005yx,Brown:2007ara} and references therein. The main results
of~\cite{Thorsson:1993bu,Brown:2005yx,Brown:2007ara} are in agreement
with those that can be obtained within a simple
relativistic mean field model~\cite{Glendenning:1997wn,Prakash:1996xs}. In all of these models the numerical value of
$\mu_e$ at a given baryon density $\rho_B$ is larger (at least a factor of 2) than the $\mu_e$ we obtain in our
calculations. Also the effective potential felt by kaons and due to interactions with $\rho$ and $\omega$ mesons is
enhanced in neutral nuclear matter, while in the model we study here it has only a mild dependence on $\rho_B$. We
guess that these two factors are the main source of difference between our results and the nuclear matter literature.
At the moment we have not yet found a solution for this problem. A first
possibility could be the hadronization of the NJL
model~\cite{Bentz:2002um}, that would allow to describe the system in terms of protons and neutrons instead of quarks.
Probably the neutralization and $\beta$-equilibration of this system would result in a larger value of $\mu_e$.
Secondly, the expansion of the effective action of the meson fields to higher orders would introduce in the theory the
direct coupling of kaons with $\omega$ and $\rho$ mesons. However a dimensional analysis shows that these contributions
are parametrically suppressed by powers of $\langle\omega_0\rangle/m_K$, thus they should not be important. In order to
verify this a direct calculation of the relevant diagrams should be performed. Thirdly, we cannot exclude that
multi-quark interactions, that are known to be useful in the stabilization of nuclear
matter~\cite{Bentz:2001vc,Huguet:2006cm,Mishustin:2003wq}, can play a
role in increasing the strength of the effective
kaon potentials. Tensor interactions might also lead to an enhancement
of the attractive interaction felt by the kaons in the medium. Last (but
not the least) there is the possibility to include a density dependence
of $G_V$ via a fit to experimental data on the $\omega$ meson
mass~\cite{Huguet:2007jc}. Some of these aspects are under current investigation.

\section{Vector interaction and color superconductivity}
\begin{figure}[t!]
\begin{center}
\includegraphics[width=8cm]{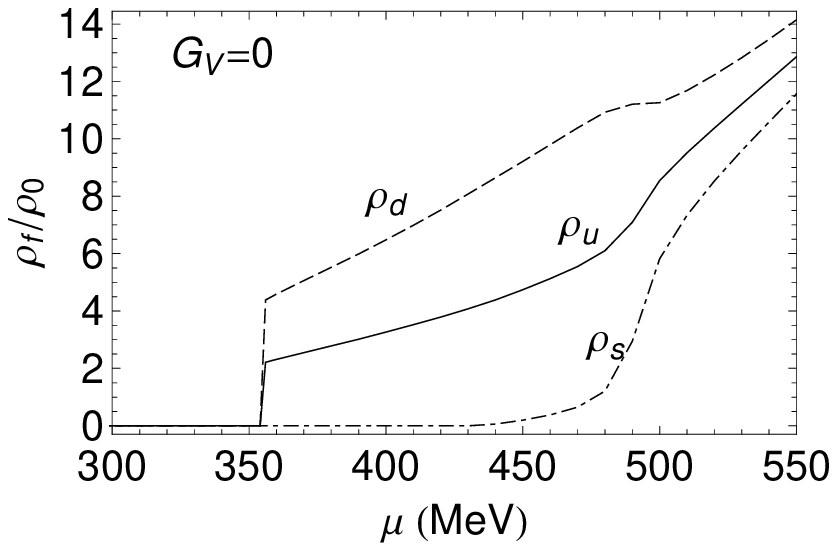}~~~~~\includegraphics[width=8cm]{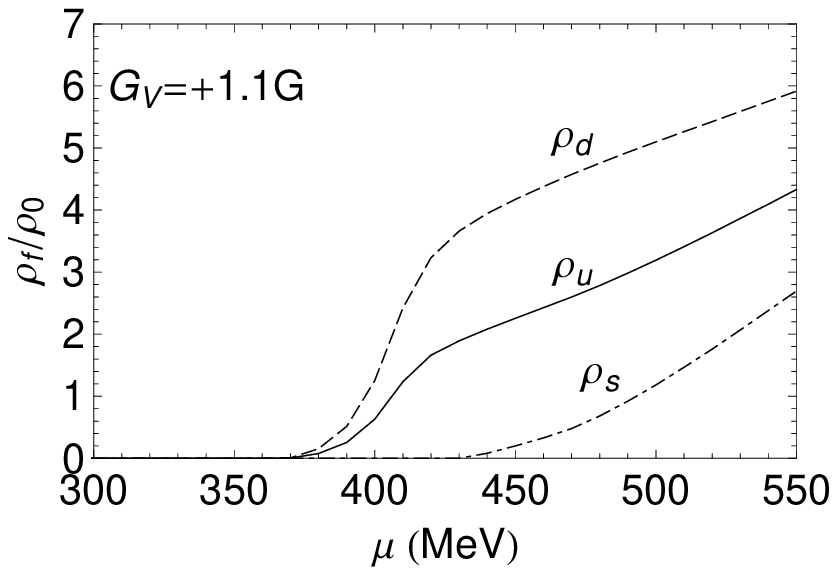}
\end{center}
\caption{\label{Fig:quarkdensities} Quark number densities against $\mu$ for $r_V = 0$ (left panel) and $r_V = +1.1$
(right panel).}
\end{figure}

We briefly comment on the effect that our results might have on
color
superconductivity~\cite{Bailin:1983bm,Alford:1997zt,Rapp:1997zu,Alford:1998mk}.
At large baryon density color superconductive phases, neglected in this
paper for simplicity, could play a relevant role in the determination of
the ground state. In the two flavor case the interactions in the
quark-antiquark and diquark channels have been considered for example in
Ref~\cite{Kashiwa:2007pc}. In order to understand how vector interaction
can interplay with superconductivity in the three flavor model under
investigation here, we compare the various quark densities at $\mu=550$
MeV for $r_V = 0$ and $r_V = 1.1$, see Fig.~\ref{Fig:quarkdensities}.

In the case $r_V = 0$ we find $\rho_u \approx 12.9 \rho_0$, $\rho_d \approx 14.2 \rho_0$, $\rho_s \approx 11.6 \rho_0$;
the in-medium quark masses ar $M_u \approx M_d \approx 7$ MeV and $M_s \approx 197$ MeV; the Fermi momenta of the
different flavors are $p_F^u \approx 538$ MeV, $p_F^d \approx 556$ MeV
and $p_F^s \approx 520$ MeV.
It is well established that in these condition quarks condense to the CFL
state~\cite{Ruester:2005jc,Abuki:2004zk,Blaschke:2005uj}. On the
other hand for the case $r_V = 1.1$ we find $\rho_u \approx 4.35\rho_0$, $\rho_d \approx 5.95 \rho_0$, $\rho_s \approx
2.71 \rho_0$; the in-medium quark masses ar $M_u \approx M_d \approx 20$ MeV and $M_s \approx 390$ MeV; the Fermi
momenta of the different flavors are $p_F^u \approx 374$ MeV, $p_F^d \approx 415$ MeV and $p_F^s \approx 320$ MeV. It
is not clear in this case which superconductive state could be favored, if any, since a self consistent calculation in
color superconductivity that keeps into account the vector interaction is missing in the three flavor case.

If we suppose that color superconductivity sets in, in order to guess which state could be a good candidate in these
conditions we compute the ratio $M_s^2/\mu$. We find $M_s^2/\mu \approx 276$ MeV. In these conditions, since $p_F^u -
p_F^s \approx p_F^d - p_F^u \approx 2(p_F^d - p_F^s) \approx 50$ MeV, probably the three flavor crystalline LOFF state
is the best candidate in the weak coupling
regime~\cite{Ippolito:2007uz,Casalbuoni:2005zp,Mannarelli:2006fy,Rajagopal:2006ig}.
If the diquark coupling is high enough at this value of the chemical
potential then a 2SC phase could be the ground state. Beside this there
is probably room for spin one condensates. We also guess that because of
the smooth crossover (see right panel of Fig.~\ref{Fig:quarkdensities})
and of the non trivial coupling between the diquark and meson
excitations a coexistence region might be created, as already discussed
in Ref.~\cite{Kashiwa:2007pc,Zablocki:2008sj}. Of course these are just
some of the possibilities, that should be either supported or not by a
dynamical calculation. We finally notice that the our reasonings are not
beyond the capabilities of the model even if we take $\mu = 550$ MeV
since the Fermi momenta in the case $r_V = 1.1$ are quite small compared
to the ultraviolet cutoff $\Lambda$ (they are smaller than the vacuum
kaon mass).

\section{Conclusions}
We have explored the consequence of the vector interaction in $\beta$-equilibrated and neutral three flavor quark
matter at finite density. Neutrality and $\beta$-equilibrium are required to reproduce the conditions that coulb be
realized in cold neutron stars. The extension of the NJL model to the VENJL one has not only an academic interest. As a
matter of fact, in view of vector manifestation the vector mesons can play a relevant role in the restoration of chiral
symmetry at large baryon density~\cite{Harada:2003jx,Harada:2000kb}. Moreover, if we wish to describe the intermediate
baryon density region of the NJL phase diagram in terms of a bosonized (and eventually hadronized) action, we need to
take into account the effects of $\omega$ and $\rho$ mesons exchange to be more realistic.

Even if it is possible to choose the value of the vector coupling $G_V$
in the vacuum, it is not clear what is its value in the medium. For this
reason we have fixed the scalar coupling $G$ and treated the ratio $r_V
= G_V/G$ as a free parameter. We have found the interesting result that
there exists a critical value of $r_V \approx 0.5$ above which the
approximate chiral restoration becomes a crossover (at $r_V = 0$ the NJL
model predicts a first order transition). This result is summarized in
Fig.~\ref{Fig:MuMu}. Our result is in agreement with Klimt, Lutz and
Weise~\cite{Klimt:1990ws} and with Fukushima~\cite{Fukushima:2008wg},
where neutrality and $\beta$-equilibrium were not required.

We have also started a systematic study of the meson energies as a function of the baryon density and on the influence
of the vector coupling, focusing in this explorative work on the pseudoscalar channels and leaving a more complete
study to a next project. In particular our findings on the $K^-$ energy show that kaon condensation is quite hard to be
realized within the present version of the model. This is disagreement with other
results~\cite{Thorsson:1993bu,Brown:2005yx,Brown:2007ara,Glendenning:1997wn,Prakash:1996xs}, but we argue that the
source of the disagreement is mainly the still poor description of the meson-quark interactions in matter at densities
of the order of few times $\rho_0$ that we have within the VENJL model. We have itemized some improvements that might
lead to a better description of matter in the intermediate density region, and some of the aforementioned research
lines are currently under investigation.

\acknowledgments We acknowledge P. Colangelo for a careful reading of the manuscript and for interesting discussions,
and M. V. Carlucci for her kind assistance in the preparation of some of the figures. We also acknowledge discussions
with C. Manuel on the topics discussed in this paper. The work of H. A. was supported by the Alexander von Humboldt
Foundation.

\end{document}